\documentclass[twocolumn]{aastex631}
\usepackage[T1]{fontenc}
\usepackage{ae,aecompl}
\usepackage{graphicx}	% Including figure files
\usepackage{amsmath}	% Advanced maths commands
\usepackage{amssymb}	% Extra maths symbols
\usepackage{multirow}
\newcommand{\be}{\begin{equation}}
	\newcommand{\ee}{\end{equation}}
\newcommand{\bq}{\begin{eqnarray}}
	\newcommand{\eq}{\end{eqnarray}}
\begin{document}
 
%\linenumbers  
 
\title { Redshift drift effect through the observation of HI 21cm signal with SKA}

 \author  {Jiangang Kang}
 \affiliation{Institute for Frontiers in Astronomy and Astrophysics, Beĳing Normal University, Beĳing 102206, China}
 \affiliation{
 	School of Physics and Astronomy, Beijing Normal University,  Beijing 100875, China}

 \author {Tong-Jie Zhang} \thanks{Email: tjzhang@bnu.edu.cn}
 
 \affiliation{Institute for Frontiers in Astronomy and Astrophysics, Beĳing Normal University, Beĳing 102206, China}
 \affiliation{	School of Physics and Astronomy, Beijing Normal University,  Beijing 100875, China}

  \author {Peng He}
 
 \affiliation{Burerau of Frontier Science and Education, Chinese Academy of Sciences, Beijing 100190}

 \author[0009-0003-6250-1396]{Ming Zhu}
 \affiliation{ National Astronomical Observatories, Chinese Academy of
 	Sciences,Beijing  100101, China}

\begin{abstract} 
This study presents the findings of using the Square Kilometre Array (SKA) telescope to measure redshift drift via the HI 21cm signal, employing semi-annual observational interval within redshift around z $\sim$ 1 with main goal is to directly gauge the universe's expansion acceleration rate with millimeter-per-second (mm/s) precision. The SKA can detect over a billion HI 21cm emissions from individual galaxies  to the redshift  z $\sim$ 2 and thousands of absorption lines from Damped Lyman-alpha (DLA) systems against bright quasars to the redshift z $\sim$ 13, with the sensitivity limit of 100 mJy.
By utilizing SKA's high spectral resolution settings (0.001, 0.002, 0.005, 0.01 Hz) to detect redshift drift, particularly focusing on the 0.001 and 0.002 Hz configuration, one aims to achieve the necessary mm/s in precision  measurement by the 0.5-year observation period.
The velocity drift rate, crucially determined by the two operational regimes within 0.01 to 0.21 mm/s and 0.031 to 0.17 mm/s, respectively, exceeds the theoretical accuracy limit of 1.28 mm/s. The analysis  thoroughly restricts cosmological parameters  related to dark energy using the Sandage-Loeb (SL) signal from the HI 21cm emission and absorption lines. It estimates $\rm H_0$ of about 70 km/s/Mpc, $\rm \Omega_m$ near 0.3, with w close to -1, $\rm w_0$ around -1, and $\rm w_a$ approaching -0.1. These results strongly endorse the SL effect as an effective method for confirming cosmic acceleration and exploring the dark sector in real-time cosmology with the SKA.
\end{abstract}
 
\section{Introduction}\label{sec:intro}
The cosmological phenomenon of redshift drift, an observationally model-independent diagnostic, also referred to as the Sandage-Loeb (SL) effect, signifies the temporal redshift fluctuations of extragalactic astrophysical sources. These fluctuations are predominantly scrutinized through detailed spectral analysis of the Lyman-$\alpha$ forest or the 21cm hyperfine transition line absorption features in the spectra of intense radio-loud quasars\citep{1962ApJ...136..319S,1998ApJ...499L.111L,2007MNRAS.382.1623B,annurev:/content/journals/10.1146/annurev-nucl-010709-151330,2015APh....62..195K}. The cosmological redshift drift serves as a pivotal diagnostic tool for examining the universe's accelerated expansion, a phenomenon ascribed to the enigmatic dark energy. Characterizing the attributes of dark energy constitutes one of the paramount challenges in contemporary cosmology\citep{1998AJ....116.1009R,1999PhRvL..83..670P,Quercellini_2012}.

Mapping the trajectory of cosmic evolution and elucidating the properties of dark energy \citep{2012PhRvD..86l3001M, 2016EPJC...76..163G, 2019MNRAS.488.3607A, 2007MNRAS.382.1623B, 2020MNRAS.492.2044C, 2021PhRvD.103h1302H} or alternative cosmological constituents require the deployment of a multitude of observational methodologies. These include the photometric analysis of Type Ia supernovae (SNe Ia), the spatial characterization of acoustic oscillations in the Cosmic Microwave Background (CMB) radiation, the identification of baryonic acoustic oscillations (BAO) within the large-scale structure power spectrum, as well as observational phenomena such as weak gravitational lensing (WL), gravitational wave (GW) detections, and other probes. Significantly, the concept of cosmic redshift drift has been proposed as an innovative technique for the direct measurement of the universe's acceleration, independent of any specific gravitational theories, spacetime curvature models, or clustering paradigms\citep{2023arXiv230204365C}. Analyzing this cosmological phenomenon through its redshift dependence can elucidate certain enigmas inherent in fundamental theoretical frameworks\citep{2021PDU....3100784K}.

To achieve precise detection of the pure radial signal over a decadal timescale, it is essential to refine the experimental spectral precision to approximately centimeters per second or better. For annual velocity variations on the order of millimeters per second, the corresponding spectral frequency shift must be detected at or below the 0.1 Hertz threshold. These stringent detection criteria are critical for an advanced single-dish radio telescope or an interferometric synthesis array.
In the context of upcoming astronomical instrumentation, the European Southern Observatory's (ESO) 40-meter class Extremely Large Telescope (ELT) will conduct cosmological redshift drift measurements through the Lyman-alpha forest, specifically targeting redshift ranges z=2 to z=5. This will involve selecting high-redshift quasi-stellar objects (QSOs) as suitable candidates for these experiments. Over a two-decade observational period, the cumulative exposure will total 4,000 hours, resulting in a radial velocity precision of 2.34 cm/s. This precision allows for the determination of the cosmological constant $\Omega_{\Lambda}$ to fall within the range [0.42,0.74] with 95\% confidence, assuming a spatially flat Universe and after marginalization over $\rm H_0$\citep{2008MNRAS.386.1192L}.

Promisingly, construction of the initial phase of the Square Kilometre Array (SKA1) commenced in 2021, strategically situated in the regions of South Africa and Australia. This phase accounts for roughly 10\% of the projected full-scale deployment of the subsequent Square Kilometre Array (SKA2). Upon its completion, the SKA is expected to be the foremost radio telescope, eclipsing all preceding instruments in capability, marking a significant technological leap a decade post the initiation of SKA1\citep{lazio2009square,rawlings2011square,2019clrp.2020...46S,2020PASA...37....7S,2019arXiv191212699B,Moresco_2022,Pritchard_2012}. The SKA will facilitate real-time cosmological investigations by detecting the 21-centimeter emission from neutral hydrogen (HI) in galaxies within the redshift range of $0 < z < 1$. Particularly, the selection of face-on galaxies emitting a distinct single Gaussian spectral profile with a higher signal-to-noise ratio (S/N) will ensure the precision of the measured signal. 
In addition to investigating DLA systems through their absorption signatures observed against optimal radio backgrounds\citep{2004NewAR..48.1259K,2015aska.confE..27K,2020EPJC...80..304L,2023MNRAS.518.2853R,2011arXiv1105.5953R,2015aska.confE.167S,2019MNRAS.488.3607A,2019arXiv190704495B,2015aska.confE.134M,Moresco_2022}, the experiment's advanced capabilities and increased sensitivity will substantially reduce the overall error of this signal. This will enable the survey to sample billions of HI galaxies at redshift z $<$ 2\citep{2015aska.confE..27K,2015aska.confE.167S} and detect over 20,000 DLA systems at a redshift of approximately 2, given the 100 mJy flux threshold of the redshifted HI (Neutral Hydrogen) 21 cm absorption line\citep{2004NewAR..48.1259K}. The Lyman-$\alpha$ forest at redshift z $<$ 1.65 is intrinsically incapable of penetrating the Earth's atmosphere. However, at such redshifts, the Universe indicates the onset of or subsequent accelerated expansion, necessitating alternative observational methodologies to elucidate these anomalies. The presence of copious neutral hydrogen (HI) in the galactic peripheries results in the emission of 21cm radiation, which remains unaffected by atmospheric interference. The dense, cold HI gas clouds within the intergalactic medium can absorb the redshifted 21cm line emitted by background sources. Consequently, the Square Kilometre Array (SKA) can leverage its superior spectral resolution and sensitivity to probe cosmic acceleration via HI 21cm observations. This facilitates the investigation of dark energy evolution and the validation of cosmological models or constraints on parameter space.
  
The 21-centimeter line, alongside Lyman-$\alpha$ radiation, emitted by neutral hydrogen atoms (HI) that are ubiquitous in both the interstellar medium (ISM) and intergalactic medium (IGM), constitutes a prolific tool for the astronomical community. This emission line permits comprehensive analyses of the neutral hydrogen gas properties within galaxies, including its spatial distribution, mass, thermodynamic state, and kinematic characteristics. Moreover, this spectral feature allows astronomers to trace the evolutionary stages of galaxies and investigate cosmological phenomena through redshift measurements. The forthcoming HI survey by the Square Kilometre Array (SKA) is projected to significantly enhance the cosmological volume probed by the 21cm emission, enabling the resolution of one billion galaxies at a redshift of approximately z $\sim$ 2, thereby extending our comprehension of the Universe. Operating in a low spectral resolution mode, the SKA is anticipated to catalog approximately $10^7$ galaxies within a redshift range of 0.2 to 1.  The observation of the HI 21cm emission from extragalactic sources at considerable redshifts ($z > 0.1$) presents significant challenges due to the inherently weak signal and the current sensitivity limitations of radio telescopes \citep{2022PASA...39...10A}. Nevertheless, Damped Lyman Alpha (DLA) systems, characterized by neutral hydrogen column densities exceeding $\rm 2 \times 10^{20}/cm^2$, have traditionally been regarded as the principal progenitors of spiral galaxies and the primary reservoirs of cosmic neutral gas mass at high redshift (z$\sim$ 3). This correlation is partially substantiated by the presence of low-ionization MgII absorption lines, which become observable in the ground-based optical range upon redshifting. It should be noted, however, that the galaxies associated with DLAs are not exclusively spiral galaxies; they also encompass dwarf galaxies and low surface brightness galaxies at lower redshifts \citep{1988qsal.proc..297W,2001A&A...369...42K,Kanekar_2001,2015A&A...575A..44G,1998AJ....116...26L,2001A&A...373..394K,2015aska.confE.134M}.
When radio galaxies or luminous quasars interact with dense HI gas clouds, significant self-shielding occurs at a critical column density, leading to the manifestation of a distinctive absorption feature at 1420.4 MHz within the clouds' continuum spectrum. These clouds typically reside in two categorically stable environments: the cold neutral medium (CNM), with temperatures spanning from 80 to 1000 K, and the warm neutral medium (WNM), with temperatures ranging from 5000 to 8000 K. The resultant spin temperature is derived as a weighted harmonic mean of the temperatures of the various phases within a multi-phase medium, dictated by the observed column density\citep{2001A&A...373..394K,Gupta_2013,2022PASA...39...10A}. Employing the HI 21cm absorption spectra as a diagnostic tool for probing cosmic expansion offers a promising trajectory for contemporary cosmology, owing to its resilience against atmospheric disturbance and insensitivity to redshift effects, thereby facilitating ground-based telescopic observations \citep{Yu_2014,Darling_2012,Jiao_2020,2022PDU....3701088L,2004NewAR..48.1259K,2023arXiv230808851K,2023EPJC...83..735C,2015mgm..conf.1590M,2019arXiv190704495B,Cooke_2019,2023MNRAS.518.2853R,2021arXiv211012242M,2020MNRAS.492.2044C,2008MNRAS.386.1192L,2021MNRAS.508L..53E}. 
  	
  	The primary aim of this investigation is to estimate the accuracy of determining the redshift drift rate as a metric for measuring cosmic expansion. This is accomplished through the utilization of the Square Kilometre Array (SKA) to observe the HI 21cm spectral emission and absorption over an observational timeframe of $\Delta t = 0.5$ year. The study subsequently strives to enhance the precision of cosmological parameters within the $\Lambda$CDM, wCDM, and CPL frameworks at an approximate redshift of z $\sim$ 1. The manuscript is structured as follows: Section \ref{sec:intro} delivers an extensive overview of the redshift drift phenomenon, characteristics of the HI 21cm spectral line, and the observational methodologies employed by the SKA; Section \ref{sec:acc} expounds the theoretical basis of the Sandage-Loeb (SL) test, cosmic acceleration models, and summarizes the systematic effects and target selection criteria in section \ref{sys}. Section \ref{res} presents the empirical results derived from the constraints imposed by HI 21cm data on cosmological parameters. Finally, Section \ref{cons} encapsulates the significant conclusions of the research.
\section{redshift drift effect model}\label{sec:acc}
The cosmological redshift drift represents a crucial metric for assessing the spacetime expansion within the context of observational cosmology. This phenomenon can be theoretically described as follows:
\begin{equation}
	\rm \dot{z}_s  =\frac{\Delta z}{\Delta t} = H_0(1+z) -H(z),
\end{equation}
Within the specified cosmological paradigm, the Hubble parameter, H(z), denotes the Universe's expansion rate at a given redshift, z. This rate is quantitatively defined as the ratio of the derivative of the scale factor with respect to time, $\dot{a}$, to the scale factor, a. H(z) encapsulates the dynamics inherent to the selected cosmological model, while $\rm H_0$ represents the Hubble constant, indicative of the present expansion rate. The observable signal is frequently articulated in terms of the spectroscopic radial velocity shift, $\rm \Delta v$, mathematically described by $\rm \Delta v = c \Delta{z_s}/(1+z_s)$, where c denotes the speed of light and $\Delta{z_s}$ signifies the variation in spectroscopic redshift $z_s$:
\begin{equation}\label{eq:v}
	\rm \dot{v} = \frac{\Delta v}{ \Delta t} =c\frac{H_0}{1+z_s}\left[1+z_s -\frac{H(z_s)}{H_0} \right] ,
\end{equation}

Similarly, the frequency shift over an observation interval $\Delta t$ for a co-moving source, as derived from $\nu_1 = \nu_{21} / (1+z)$ and $\nu_2 = \nu_{21} / (1+z+\Delta z)$, can be expressed as: 
\begin{equation}
	\rm \Delta \nu = \nu_2 -\nu_1 \approx  -\nu_{21} \frac{\Delta z}{(1+z)^2},
\end{equation}
In this context, $\rm\nu_{21} = 1420.405751768$ MHz represents the rest frame frequency of the HI 21 cm absorption line, with $z$ indicating the initially determined redshift, and $\nu_2$ and $\nu_1$ denoting the first and second observed frequencies, respectively.

\begin{figure}
	\begin{center}
		\includegraphics[width=0.5\textwidth]{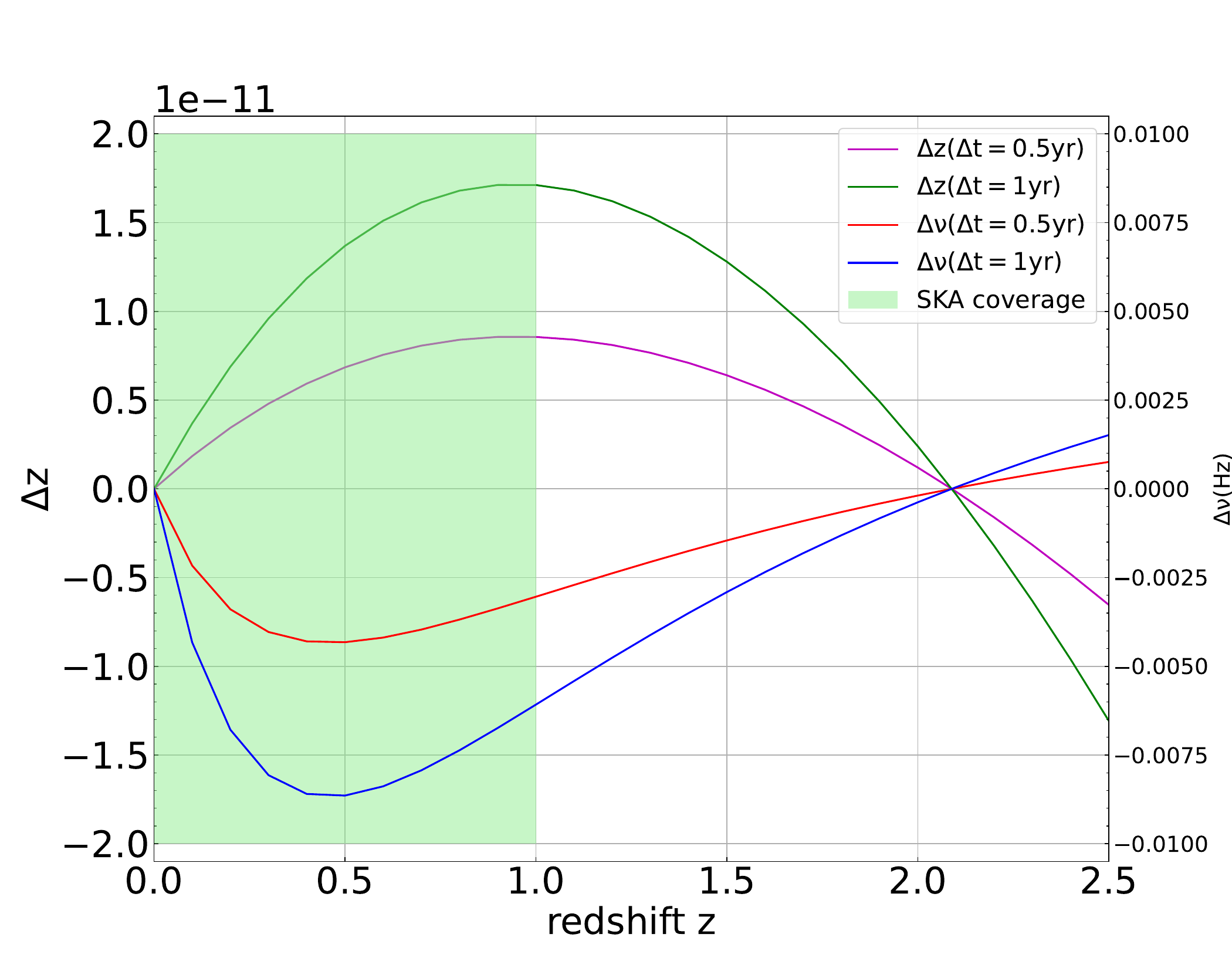}
		\caption{Theoretical variations of redshift drift $\Delta{z}$ (left) and frequency resolution $\Delta\nu$ (right) as functions of redshift, considering an observational duration of 0.5 year and 1 year within the framework of $\rm \Lambda CDM$ cosmology (P18). The light green region delineates the redshift range accessible to SKA observations.\label{figs:zh}}
	\end{center}
\end{figure}

It is well-known that the universe is experiencing a phase of acceleration at a redshift close to z=0.67, having transitioned from a previous era of deceleration. Consequently, the velocity drift exhibits positive values at lower redshifts, inversely shifting to negative values as redshift increases\citep{2021arXiv211012242M,2023arXiv230808851K}. The limiting velocity drift resulting from a time interval of $\Delta t$ = 0.5 year with the full SKA configuration is approximately 1.28 mm/s within the redshift range z$\sim$ 1. It is anticipated that the experiment will be conducted using high spectral resolution datasets ($\Delta \nu$ = 0.001, 0.002, 0.005, 0.01 Hz) to further investigate this phenomenon\citep{2015aska.confE..27K,2019MNRAS.488.3607A,2020EPJC...80..304L}. It is crucial to acknowledge that, regarding the feasibility of utilizing the HI 21cm signal, only spectral resolutions of $\Delta \nu$ = 0.001 Hz and 0.002 Hz can achieve the necessary accuracy throughout the entire redshift range of z=0 to z=1. Figure \ref{figs:zh} delineates the necessary precision for redshift measurements and the detection of frequency shifts over observational intervals of 0.5 and 1 year, respectively, as forecasted by the standard $\Lambda$CDM cosmological model based on the Planck 2018 data release (P18)\citep{2020A&A...641A...6P}. The graph indicates that the redshift measurement precision $\Delta z$ reaches an average magnitude of approximately $10^{-11}$ on the left ordinate, while the frequency resolution $\Delta\nu$ on the right ordinate achieves a sub-hertz level, specifically below 0.1Hz, over observational periods of either half a year or a full year. This enhanced resolution, enabled by the SKA and reduced observation intervals, imposes the spectral discrimination and requisite stability for the detection of the SL signal. 

\begin{figure}
	\begin{center}
	\includegraphics[width=0.5\textwidth]{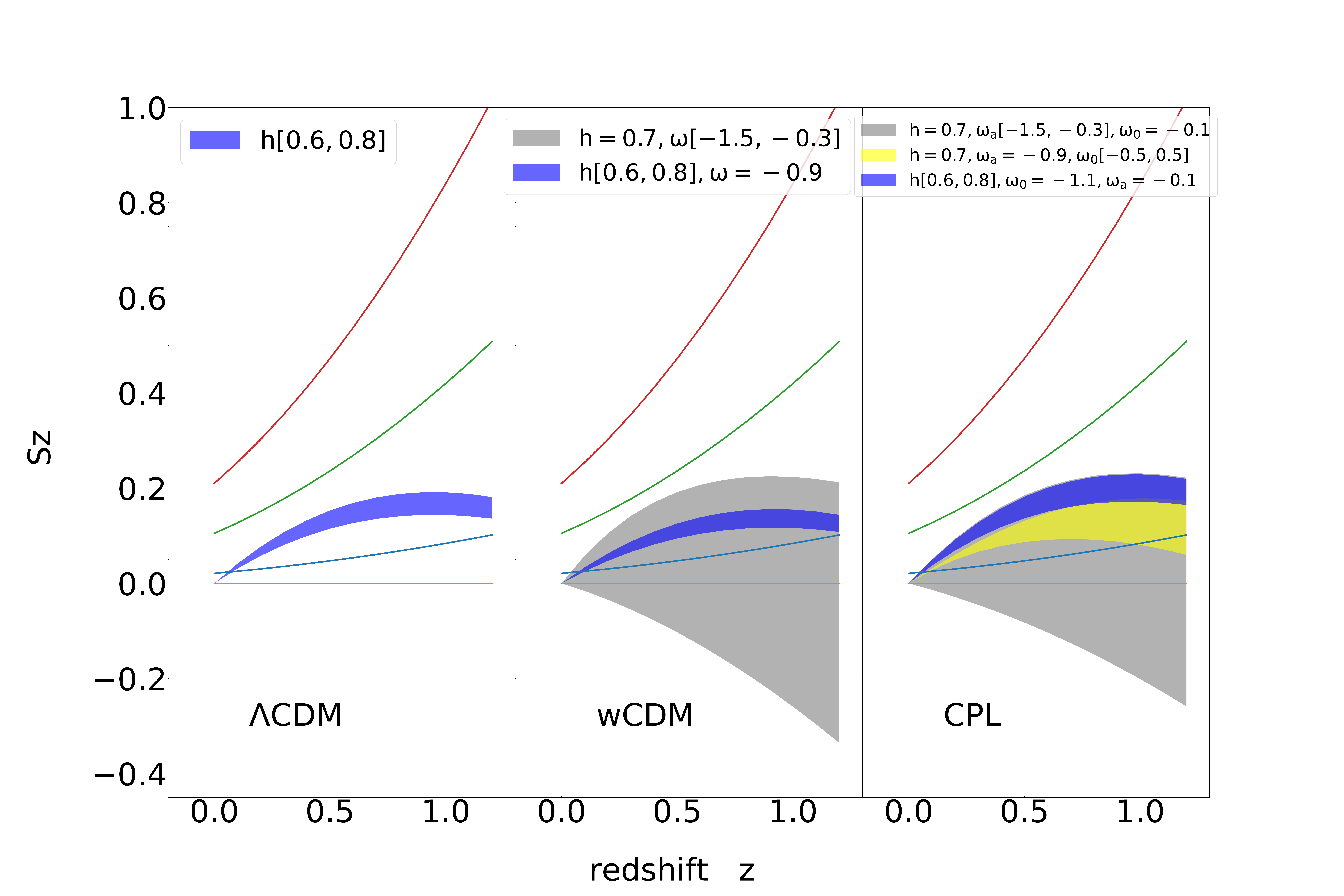}
		\caption{The variation of the dimensionless redshift drift with different cosmological models and redshifts is depicted, where the colored lines from bottom to top  denote the spectral resolution capabilities of the SKA (0.001, 0.002, 0.005, 0.01 Hz), and the shaded regions in various colors indicate the ranges of specific parameter variations.}\label{figs:sz}
	\end{center}	 
\end{figure}
\begin{figure}[!tph]
	\begin{center}
		\includegraphics[width=0.5\textwidth]{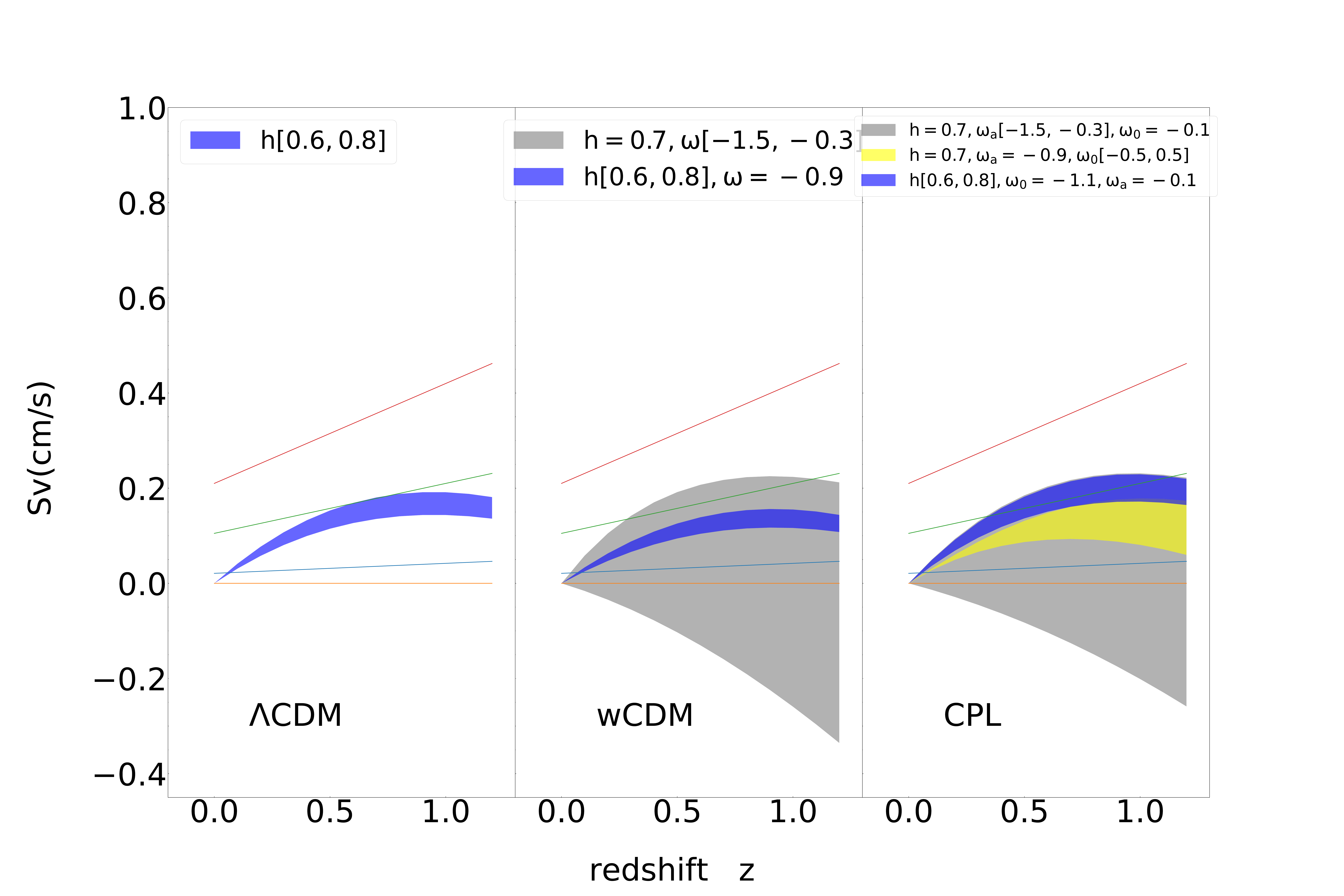}
		\caption {The variation in dimensionless velocity drift as a function of specific models and redshift is illustrated by four distinct lines from bottom to top, corresponding to the spectral resolution capabilities of the SKA (0.001, 0.002, 0.005, 0.01 Hz). The shaded regions in different colors represent the range of specific parameters variations.}\label{figs:sv}
	\end{center}	 
\end{figure}
Considering the spectral resolution capabilities of the SKA at 0.001, 0.002, 0.005, and 0.01 Hz, we rigorously define the quantification of the dimensionless redshift drift, denoted as $\rm S_z$, and the velocity drift, indicated as $\rm S_v$, following the definitions established in prior research\citep{2019MNRAS.488.3607A,2021arXiv211012242M}.
\begin{equation}\label{eq:sz}
	\rm Sz = \frac{1}{ H_{100}}  \frac{\Delta z}{\Delta t}=h\left[1+z-E(z) \right],
\end{equation}
\begin{equation}\label{eq:sv}
	\rm Sv = \Delta v= kh \left[1+\frac{E(z)}{1+z}\right]
\end{equation}
In this context, $\rm H_0 = hH_{100}$ where $\rm H_{100}$ represents 100 km/s/Mpc, and $\rm E(z) = H(z) / H_0$. The parameter $\rm k = cH_{100}\Delta t$. Figures \ref{figs:sz} and \ref{figs:sv} illustrate the two variables within the $\Lambda$CDM, wCDM, and CPL cosmological models. The two graphical depictions demonstrate that, within the intrinsic uncertainties of the Hubble constant $H_0$ ranging from 60 to 80 km/s/Mpc, spectral resolutions of 0.001Hz and 0.002Hz are plausible scenarios for the unequivocal detection of signals spanning the entire redshift range from 0 to 1. This achieves observational precision levels of 0.21 to 0.42 mm/s and 0.42 to 0.84 mm/s, respectively, and accommodates fluctuations in the dark energy equation of state parameters $w$, $w_0$, and $w_a$. The corresponding color-coded maps in blue, gray, and yellow illustrate these specific parameters variations as indicated in the legend, while the numerical values associated with these specific parameters variations are detailed in Table \ref{tabs:sensi}.
\begin{table} 
	\centering
	\begin{tabular}{c|c|ccc}
		\hline
		\hline
		
		quantity& mode &$\rm \Lambda CDM$ & $\rm wCDM $&$\rm CPL$\\                     
		
		\hline
		\multirow{2}{*}{$\rm Sz$ }& th& [0,0.3] &  [-0.4,0.4] &  [-0.35,0.35]  \\
		\cline{2-5}                             
		& obs&\multicolumn{3}{c}{[0,0.65]}       \\
		
		\hline
		\multirow{2}{*}{$\rm Sv$ } &th & [0.0,0.15] & [-0.2,0.2] & [-0.14,0.18]  \\
		
		\cline{2-5}  
		& obs&\multicolumn{3}{c}{[0.02,0.48]}    \\
		
		\hline
		
	\end{tabular}\caption{The magnitude of the dimensionless redshift drift and velocity drift derived from theoretical predictions and observational data, as depicted in Figures \ref{figs:sz} and \ref{figs:sv}, under three distinct cosmological models, are represented by "th" and "obs" respectively. \label{tabs:sensi}}
\end{table}

\section{Systematic effects and Target selection}\label{sys}

Systematic effects encompass the stability of baseline and frequency calibration, the relative angular displacement of absorbing gas with respect to background objects, variations in the illumination source's size, flux, and spectral characteristics, observer motion, peculiar velocities and accelerations, as well as gravitational accelerations within and between targeted objects. The necessary calibration stability is contingent on a local oscillator, and current radio facilities are capable of supporting this level of precision \citep{2020MNRAS.492.2044C}. If experiments are conducted in a highly non-inertial reference frame with multiple accelerations and rotations, the precision will be thoroughly addressed in the SKA era, potentially surpassing the requirements for signal detection \citep{2015aska.confE..27K,2020MNRAS.492.2044C,2015APh....62..195K,Moresco_2022,2019MNRAS.488.3607A}. Moreover, the sign of $\dot{z}$, a well-documented phenomenon at both low and high redshifts (deviating from the null value), and the directions of gravitational accelerations are stochastic and null-centered. Consequently, the cumulative impact of peculiar accelerations will be mitigated by expanding the HI 21cm sample sizes across diverse sky regions, increasing integration time, and extending the duration of the experiments \citep{Moresco_2022}. Regarding the feasibility of experiments with the SKA \citep{2015aska.confE..27K}, for extragalactic sources with redshifts exceeding 0.2, the peculiar motion in redshift space will be attenuated to $10^{-14}$, which is an order of magnitude lower than the percent-level cosmological signal and can be disregarded in forthcoming observations.

  In addition to the frequency stability and gravitational accelerations of HI 21cm signals, numerous technical constraints need thorough examination, broadly classified into three categories: Firstly, the observation will be conducted within an ultra-stable reference framework, with no temporal or positional deviations relative to the targets. Astrometric accuracy, time standards, and pointing precision must be calibrated and adjusted to an accuracy of 1 arcsecond. These inevitable interferences will be systematically addressed as data streams from spectral channels into processing pipelines. Secondly, pulsar observations provide an unbiased solution due to their precise timing capabilities, allowing for long-term stability assessments of the global SKA system by analyzing the pulse arrival times of multiple pulsars to eliminate systematic effects. Additionally, a vast number of correlator channels will be employed to comprehensively capture the neutral hydrogen signals of spiral galaxies. Determining the redshift drift requires $10^{8}$ $\sim$ $10^{9}$ correlator channels, a challenge that can be effectively managed through meticulous spectral window design and frequency standardization during data acquisition\citep{2015aska.confE..27K,2020PASA...37....7S}. 
   
  Assuming that the precise identification of targeted samples can generate sufficient radiation at any redshift to achieve reasonable error margins of $\rm\sigma_v \le 1 mm/s/year$, especially for emission-line galaxies. For HI 21cm emission, galaxies exhibiting a single Gaussian profile will be ideal targets, provided they have peak signals exceeding 100 mJy and a signal-to-noise ratio (S/N) of at least 100, due to the more stable baseline and superior sensitivity of the SKA system. Conversely, Damped Lyman Alpha (DLA) systems with multiple Gaussian components will yield lower errors for SL signals, making intervening-DLA systems the primary targets\citep{2020MNRAS.492.2044C,2023arXiv230204365C,Darling_2012,2020EPJC...80..304L,2015aska.confE.167S,2015aska.confE..17A,2022PASA...39...10A,1988qsal.proc..297W,2001A&A...369...42K,2024MNRAS.tmp.2497E,2024arXiv240806626Y,2024MNRAS.tmp.2497E,2015APh....62..195K,2023MNRAS.518.2853R,2004NewAR..48.1259K,2020PASA...37....7S,2015aska.confE.134M,2024RAA....24g5002K,2015MNRAS.450.2251Y,2022JApA...43..103D}.  
  
\section{result and discussion}\label{res}
This section elucidates the methodologies employed for estimating velocity drift and its consequential implications for cosmology. In alignment with the findings depicted in figure \ref{eq:sz}, the precision of velocity shift estimation can be achieved by leveraging the SKA's spectral capabilities of 0.001 Hz and 0.002 Hz. Two configurations are thoroughly examined in detecting the 21cm line emissions of neutral hydrogen (HI) from galactic sources, as well as the absorption features of Damped Lyman Alpha (DLA) systems, over a redshift range of $0<z<1$. Regarding different cosmic epochs, specifically the epoch of cosmic evolution with redshift $2<z<5$, the Extremely Large Telescope (ELT) serves as a complementary instrument. Furthermore, constraints on $\rm H_0$, $\rm\Omega_m$, and the dark energy equation of state parameters $\omega$, $\omega_0$, and $\omega_a$ are considered, taking into account the $\rm H_0$ uncertainty spanning from 60 to 80 km/(sMpc). The chi-squared ($\chi^2$) distribution is applied to the velocity shift ($\Delta v$), which informs the posterior distribution used to compute the likelihood functions for the parameter sets within each respective model.
\begin{equation}
	\chi^2 = \sum_{i=1}^{N}\frac{(\Delta v_{obs}- \Delta v_{th})^2}{\sigma_i^2} ,
\end{equation}
In the equation, $\Delta v_{th}$ represents the theoretical velocity as delineated by equation \ref{eq:v}, whereas $\Delta v_{obs}$ denotes the observed velocity for the i-th data point, ascertained through SKA observations with ultra-high spectral resolutions of 0.001 Hz and 0.002 Hz, as referenced in \citep{2015aska.confE..27K}. Additionally, $\rm \sigma_{i}$ signifies the standard error attributed to the i-th data point, calculated as the reciprocal of the square root of the source count ($\rm\sigma_{i}$ $\sim$ $\rm 1/\sqrt{N}$), where $\rm N$ denotes the number of galaxies or DLA systems detectable at these specified resolutions established on the SKA platform.

\begin{equation}\label{eqs:n6}
	\rm \frac{dN/dz}{deg^2} =10^{c_1}z^{c_2} exp(-c_3 z),
\end{equation}

\begin{equation}\label{eqs:n7}
	\rm b_{HI} (z) = c_4 exp(c_5 z).
\end{equation}

\begin{figure}
	\includegraphics[width=.55\textwidth]{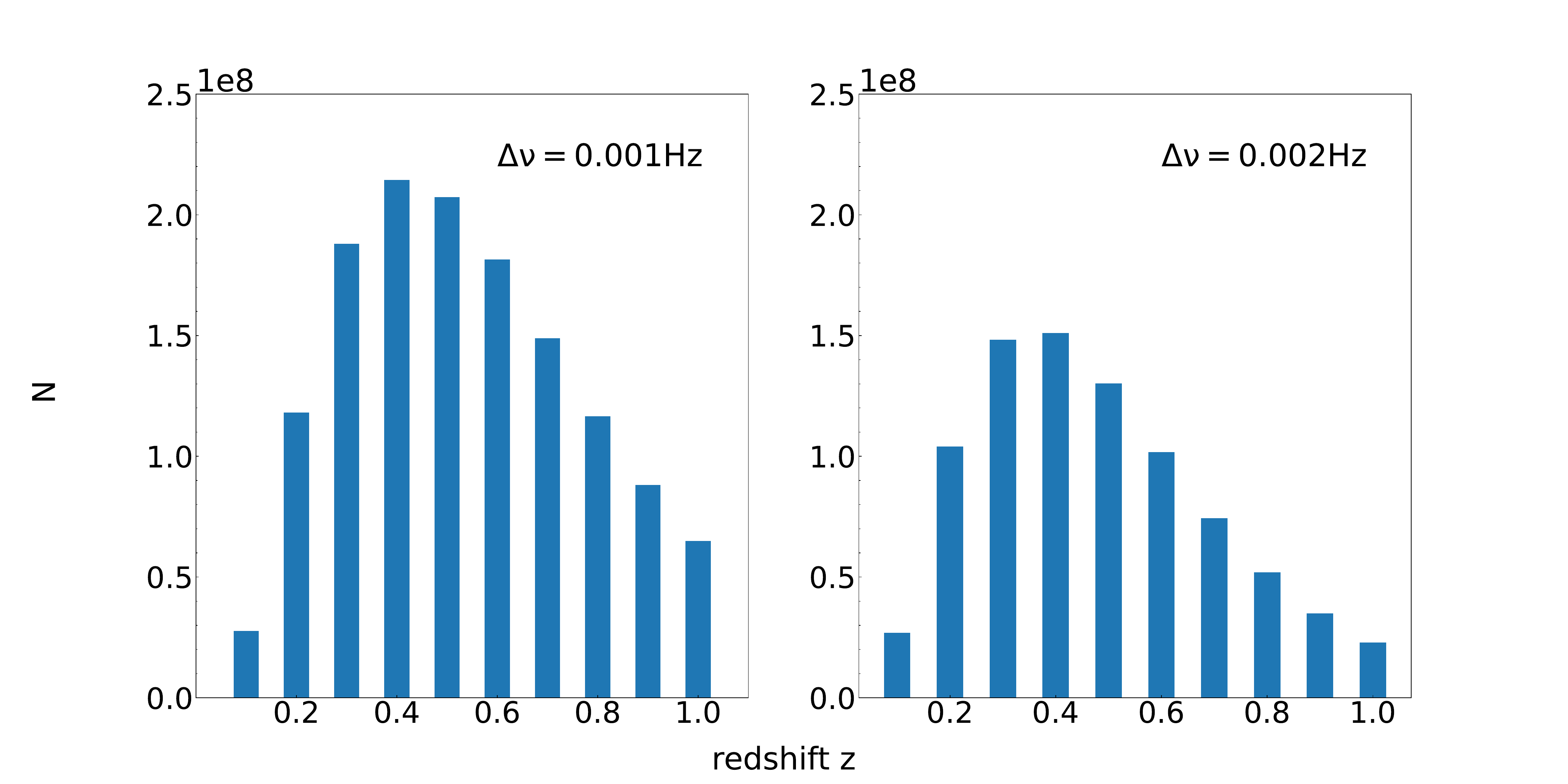}
	\caption{The redshift distribution of the number of extragalactic HI 21cm emissions detected by SKA at a spectral resolution of 0.001Hz (left) and 0.002Hz (right). \label{figs:num}}
\end{figure}
\begin{figure}
	\includegraphics[width=.5\textwidth]{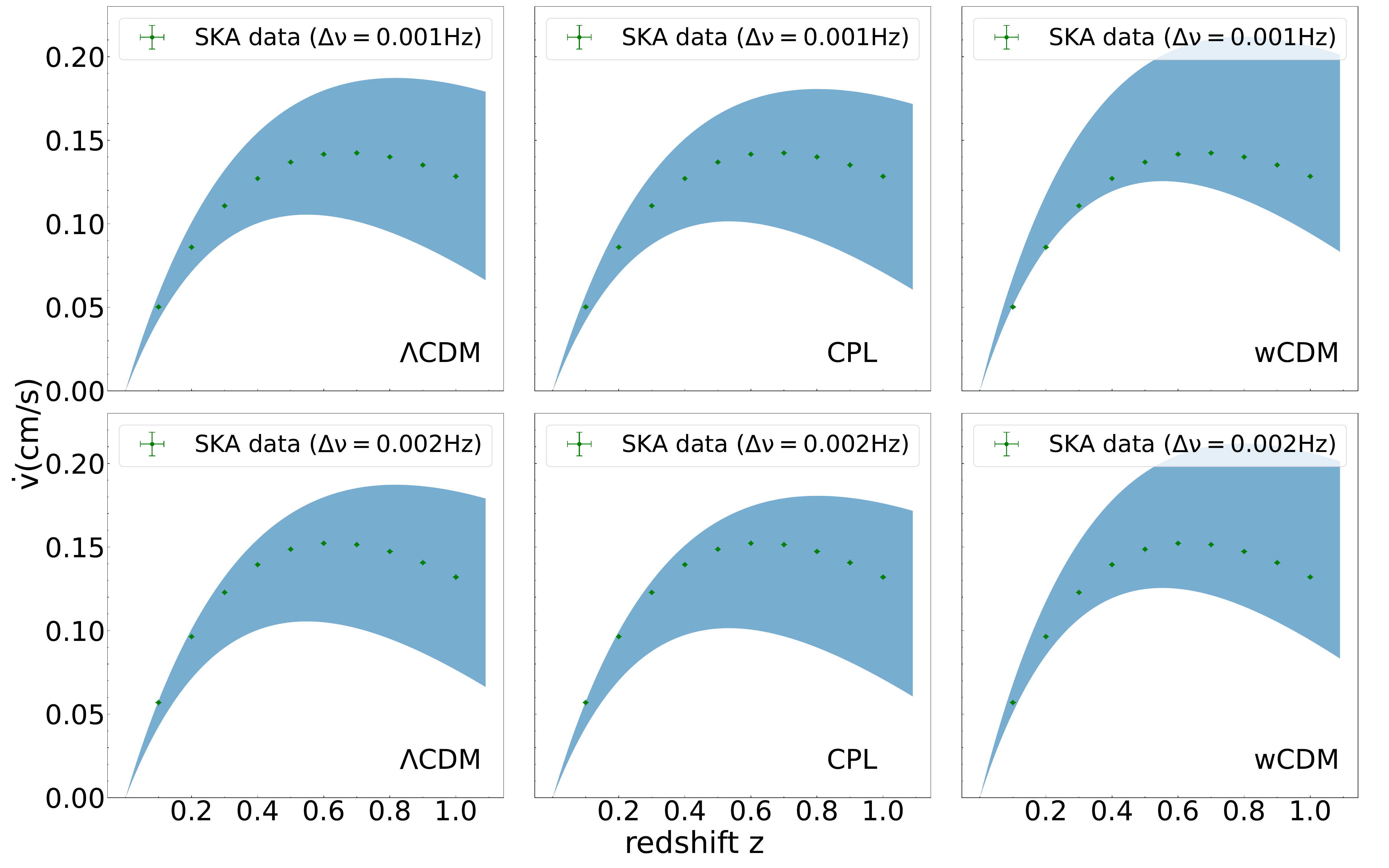}
	\caption{The light blue region in each graph represents the extent of velocity drift, taking into account the uncertainty in $\rm H_0$ ranging from 60 to 80 km/s/Mpc over a 0.5-year observation period within the redshift interval of 0 to 1. The observed redshift and velocity drift data from SKA, with spectral resolutions of 0.001 Hz (top panel) and 0.002 Hz (bottom panel), derived from the HI 21cm emission of individual galaxies, are depicted with green cross error bars. }\label{figs:em}
\end{figure}

The objective is to ascertain the velocity drift error, denoted as $\sigma_{i}$, through the utilization of the galaxies' redshift distribution. This can be accomplished by computing the detection distribution, represented by the function dN/dz, using the specific equation \ref{eqs:n6}, while the HI galaxy bias can be validated via equation \ref{eqs:n7}. The SKA conducts this analysis for each 0.1 redshift bin within the range of z=0 to z=1 \citep{2009ApJ...703.1890O,2015MNRAS.450.2251Y}. The coefficients $c_1$, $c_2$, $c_3$, $c_4$, and $c_5$ are variables associated with the respective detection limits, as indicated in table 4 of \citep{2015MNRAS.450.2251Y}. In our study, we utilize the constants $c_1$, $c_2$, $c_3$, $c_4$, and $c_5$ with values of (6.532, 1.932, 5.224, 0.530, 0.781) for the 0.001Hz scenario and (6.555, 1.932, 6.378, 0.549, 0.812) for the 0.002Hz case to determine the number of distant galaxies (N) detectable via the HI 21cm emission line.

 In Figure \ref{figs:num}, the left panel demonstrates the SKA's proficiency in identifying extragalactic distributions with an exceptional spectral resolution of 0.001 Hz. For each 0.1 increment in redshift, there is an average density of $1.36 \times 10^8$ galaxies, with a peak density of approximately $2.14\times  10^8$ galaxies at a redshift of 0.4, and a minimum of $2.8 \times  10^7$ galaxies at a redshift of 0.1. Cumulatively, at a spectral resolution of 0.001 Hz, the total number of galaxies observed up to a redshift of 1 is nearly 1.36 billion. In comparison, at a spectral resolution of 0.002 Hz, the maximum galaxy density is around $1.51 \times  10^8$ at a redshift of 0.4, with a cumulative count of approximately $8.46\times  10^8$ galaxies by a redshift of 1, and a minimum of $2.69 \times  10^7$ galaxies within the redshift bin of 0.1. Upon analyzing galaxy counts, as the redshift z exceeds 0.4, a significant reduction is observed in both scenarios, highlighting the critical importance of distribution analysis in understanding galactic evolution and their cosmological implications. A comparative assessment reveals that the scenario with a spectral resolution of 0.001 Hz significantly outperforms the 0.002 Hz resolution across all redshift bins. The maximum discrepancy between these cases is observed at a redshift of approximately 0.6, where the count difference reaches 9.87 million, and by a redshift of around 1, the total divergence amounts to $5.09 \times  10^8$. Preferentially, face-on galaxies are utilized to determine the redshift drift effect due to their prominent single Gaussian emission profiles.

Consequently, the observed velocity drifts $\rm \Delta v_{obs}$ on the SKA can be quantified under the two spectral resolution scenarios of 0.001 Hz and 0.002 Hz. This quantification depends on the assessment of the total  number of sources of HI 21cm signals, N, within each redshift bin, z, encompassing both emission and absorption line data, as well as the observational duration, $\rm\Delta T$. This relationship can be precisely modeled as the following formalism \citep{2015aska.confE..27K,2008MNRAS.386.1192L,2022MNRAS.514.5493D,2021MNRAS.508L..53E,Alves,2019arXiv190704495B}:
 \be\label{eqs:sv1}
 \rm \Delta v_{obs}=\sigma_n N^{-1/2}(1+z)^{\lambda}\rm \Delta \nu^{1/2}   \quad [ cm/s],
 \ee
 The parameter $\rm\sigma_{n}$ denotes a normalization constant. For emissions from extragalactic sources, we project an approximate count of N=$10^7$ per 0.1 redshift interval up to z=1, over an observational duration $\Delta T = 0.5$ years, employing spectral resolutions of 0.001 Hz and 0.002 Hz within the redshift span $0 \leq z \leq 1$. The parameter $\lambda$ is held constant at 1.09 for the 0.001 Hz data and 1.52 for the 0.002 Hz dataset. In terms of detection accuracy, theoretical analyses indicate that with an observation period of $\Delta T = 0.5$ years, the SKA is capable of detecting redshift drift at spectral resolutions between 0.001 and 0.01 Hz \citep{2015aska.confE..27K}. It is optimal for the data precision not to exceed 0.002 Hz as the redshift approaches unity. The normalization constant $\sigma_{n}$ ranges from 1 to 5 cm/s, varying with spectral resolution and redshift (Kang 2024, in prep), and decreases linearly in both contexts. At lower redshifts or spectral resolutions, the values are notably higher, especially when redshift exceeds 0.8 or when spectral resolution surpasses 0.01 Hz, maintaining a constant error margin of 0.25. Consequently, the substantial number of targeted sources with adequate signal-to-noise ratios (S/N >= 100) during the SKA era will guarantee the experiment's reliability.

  Figure \ref{figs:em} illustrates the velocity drift values, denoted as $\rm\dot{v}$, represented by green cross error bars, along with those derived from three distinct theoretical models that hypothesize a systematic deviation in the Hubble constant ($\rm H_0$) ranging from 60 to 80 km/s/Mpc, depicted in a shallow blue contour map. The upper panel corresponds to a frequency of 0.001 Hz, while the lower panel pertains to a frequency of 0.002 Hz. The x-axis and y-axis denote the redshift drift and the velocity drift, respectively. Through semiannual observational analysis ($\Delta t = 0.5$ year) as shown in the figure, the empirically determined velocity drift rates range from 0.05 to 0.15 cm/s, with the associated uncertainties quantified between 0.01 and 0.03 cm/s.
  
\begin{figure}
	\includegraphics[width=.5\textwidth]{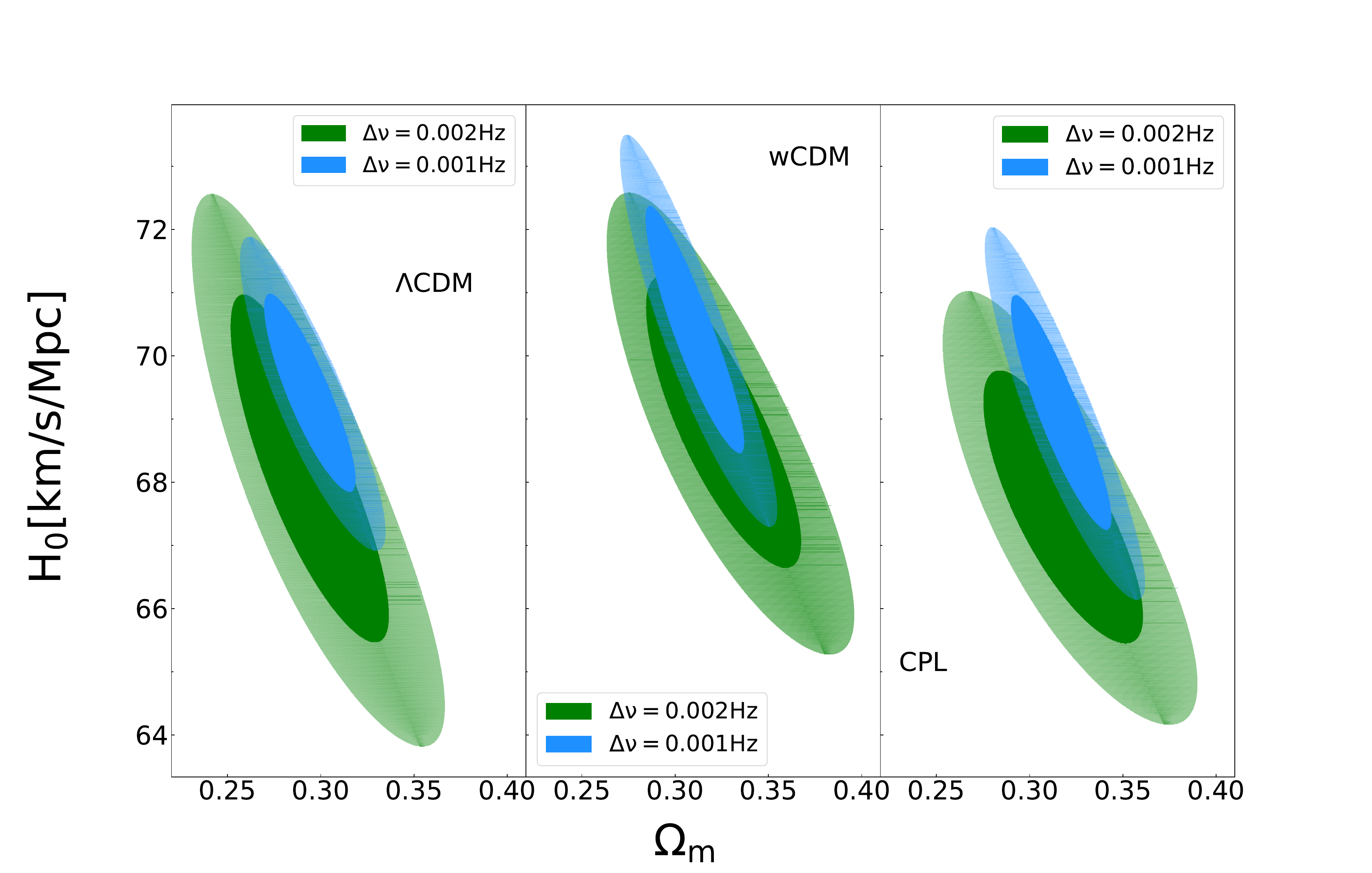}
	\caption{The constraints on the $H_0$ and $\Omega_m$ parameter space within the 1$\sigma$ and 2$\sigma$ confidence intervals, derived from the predicted data utilizing the SKA's spectral resolutions of 0.001 Hz (blue contours) and 0.002 Hz (green contours).}\label{figs:obs1}
	\end{figure}

 Regarding the recorded metrics of the two panels, the mean value derived from the 0.002Hz frequency is marginally higher, by 0.003-0.007 cm/s, compared to the 0.001Hz data, particularly at lower cosmological volumes. This degree of precision aligns perfectly with the predictions from various theoretical frameworks as outlined in Equation \ref{eq:v}. As illustrated in Figure \ref{figs:obs1}, within the $\Lambda$CDM and wCDM cosmological models, the analysis suggests a preference for a Hubble constant ($\rm H_0$) around 70 km/s/Mpc. In contrast, under the CPL parametrization, the data indicate a slightly lower $\rm H_0$ value, approximately 67 km/s/Mpc.
\begin{figure}
	\includegraphics[width=.5\textwidth]{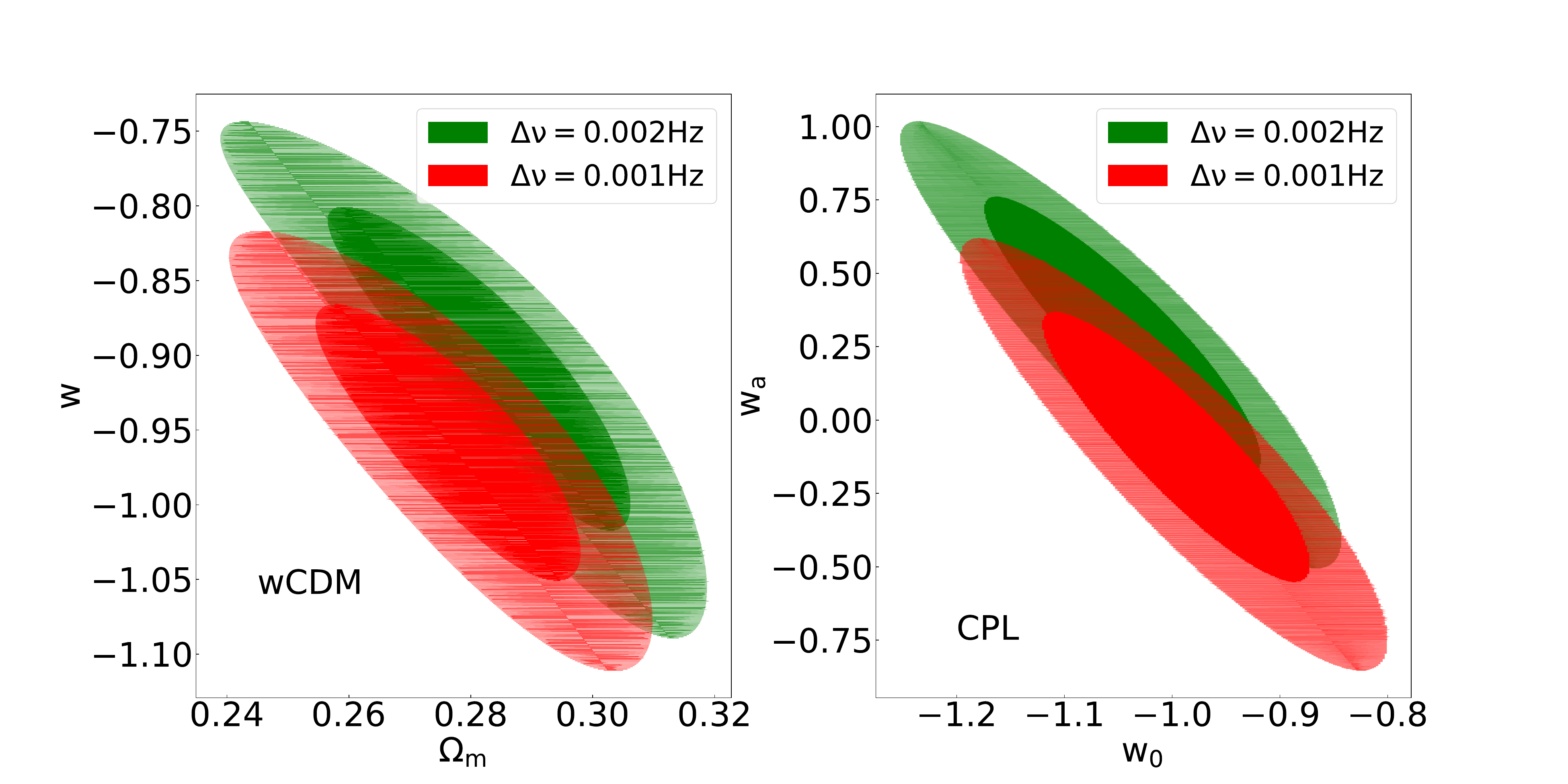}
	\caption{The constraints on $\rm w$, $\rm w_0$, and $\rm w_a$ within the 1$\sigma$ and 2$\sigma$ confidence intervals, derived from the forecasted data using the SKA spectral resolutions of 0.001 Hz (depicted by red contours) and 0.002 Hz (depicted by green contours).}\label{figs:obs2}
	
\end{figure}

\begin{table} 
	\caption{The priors applied to the free parameters in the $\Lambda$CDM, wCDM, and CPL cosmological models.}\label{tabs:prior}
	\centering
	\begin{tabular}{c|c}
		\hline

		parameter&   Proir  \\                     
		\hline
		$H_0$      &    [60,80] \\
		\hline         
		$\Omega_m$ & [0.1,0.6]\\
		\hline
		w       &   [-3,3]  \\  
		
		\hline
		$\rm w_0$&[ -3,3]  \\
		\hline
		$\rm w_a$ & [-3,3] \\
	 
		\hline
\end{tabular}
\end{table}

\begin{table*}[t] 
	\caption{The optimal parameter values and 1$\sigma$ (68.3\%) confidence intervals for constraints at 0.001Hz under three different cosmological frameworks.\label{tabs:bf1}}
	\centering
	\resizebox{\textwidth}{!}{
		\begin{tabular}{cccccc}
			\hline
			\hline
			
			model& $ H_0$ & $\Omega_m$ &w &$\rm w_0$ &$\rm w_a$ \\                     
			\hline
			$\Lambda$CDM&  $69.6^{+1.07}_{-0.94}$ &$0.311^{+0.304}_{-2.214}$ & -- & --  &--\\
			\hline         
			
			wCDM &$70.4^{+0.81}_{-0.72}$  & $0.317^{+0.087}_{-0.142}$  &   $-0.953^{+0.077}_{-0.101}$ & --  & --\\
			
			\hline

			CPL &$69.3^{+1.22}_{-1.09}$ & $0.324^{+0.284}_{-0.165}$ & --& $-0.992^{+0.118}_{-0.251}$ &$-0.124^{+0.385}_{-0.439}$\\
			\hline
			
	\end{tabular}}
\end{table*}
\begin{table*} \caption{The optimal value and 1$\sigma$ (68.3\%) confidence intervals for the 0.002Hz observational data from SKA within the framework of the three cosmological models.\label{tabs:bf2}}
	\centering
	\resizebox{\textwidth}{!}{
		\begin{tabular}{cccccc}
			\hline
			\hline

			model& $ H_0$ & $\Omega_m$ &w &$\rm w_0$ &$\rm w_a$ \\                     
			\hline
			$\Lambda$CDM&  $68.26^{+2.881}_{-3.046}$ &$0.294^{+0.047}_{-0.028}$ & -- & --  &--\\
			\hline         
			
			wCDM &$69.24^{+2.902}_{-2.397}$  & $0.321^{+0.087}_{-0.142}$  &   $-0.861^{+0.071}_{-0.153}$ & --  & --\\
			
			\hline
			CPL &$67.13^{+1.822}_{-1.453}$ & $0.328^{+0.271}_{-0.193}$ & --& $-1.03^{+0.130}_{-0.231}$ &$0.244^{+0.264}_{-0.517}$\\
			\hline
			
	\end{tabular}}
\end{table*}

In relation to the emission findings, we impose constraints on the cosmological parameters within the frameworks of the standard $\Lambda$CDM, wCDM, and CPL parametrization models by leveraging the anticipated data from the 21cm emission line. The best-fit parameter estimates for $\rm H_0$, $\Omega_m$, $w$, $w_0$, and $w_a$ are depicted along with their corresponding 68.3\% (1$\sigma$) and 95.4\% (2$\sigma$) confidence levels (CL) in Figures \ref{figs:obs1}--\ref{figs:obs2} and in Tables \ref{tabs:bf1} and \ref{tabs:bf2} for the 0.001Hz and 0.002Hz data, respectively. The prior distributions for these parameters are detailed in Table \ref{tabs:prior}. Analysis of the panels reveals that the constrained data is in close concurrence with results from other observational probes, with a significantly narrowed parameter space for all parameters, especially for $\rm H_0$ and $\Omega_m$ in the CPL model.

Regarding the findings at 0.001Hz and 0.002Hz frequencies, the Hubble constant \( H_0 \) was confined within the range of 70.4 to 67.13 km/s/Mpc. The precision in the \(\Lambda\)CDM model has been enhanced, tightening from 1.12\% to 0.8\%. For the wCDM model, the precision has increased from 1.19\% to 1.06\%, and in the CPL model, from 0.91\% to 0.84\%. Furthermore, constraining the matter density parameter \(\Omega_m\) within 0.294 to 0.328 yielded precision improvements as follows: in the \(\Lambda\)CDM model, from 1.05\% to 0.88\%; in the wCDM model, from 1.33\% to 1.08\%; and in the CPL model, from 1.52\% to 1.17\%. In the wCDM model, the accuracy of the dark energy equation of state parameter \( w \) has improved from 2.24\% to 1.65\%. In the CPL model, the precision of parameter \( w_0 \) has increased from 4.55\% to 3.66\%, and for \( w_a \), the enhancement is from 4.30\% to 3.61\%. This indicates that parameter degeneracies are effectively mitigated when integrating SKA observations with other cosmological datasets.
\begin{figure}
	\includegraphics[width=0.5\textwidth]{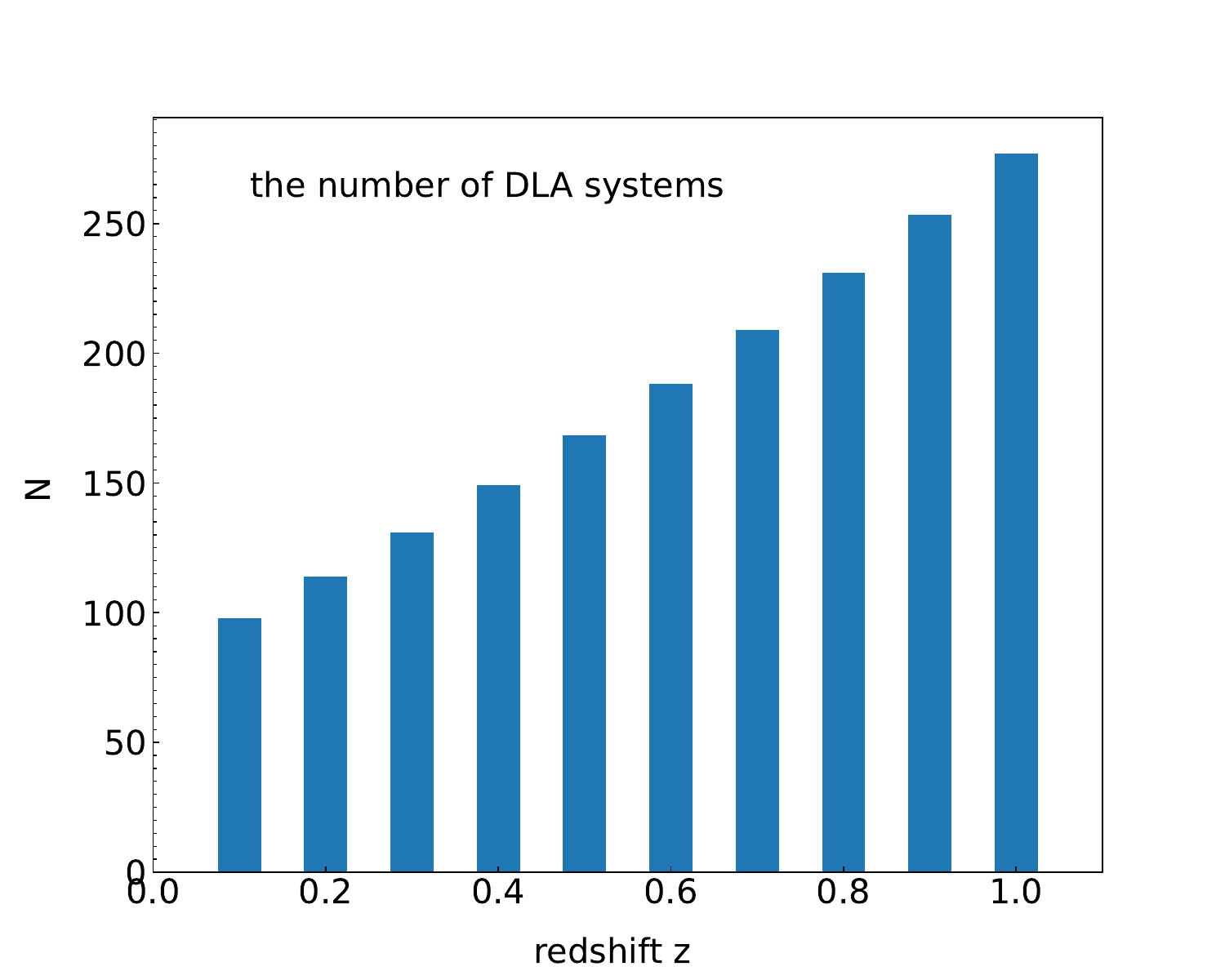}
	\caption{The census of Damped Lyman-Alpha (DLA) systems at redshift intervals from 0 to 1 identified using the Square Kilometre Array (SKA). }\label{figs:ab}
	
\end{figure}

\begin{figure}
	\includegraphics[width=.5\textwidth]{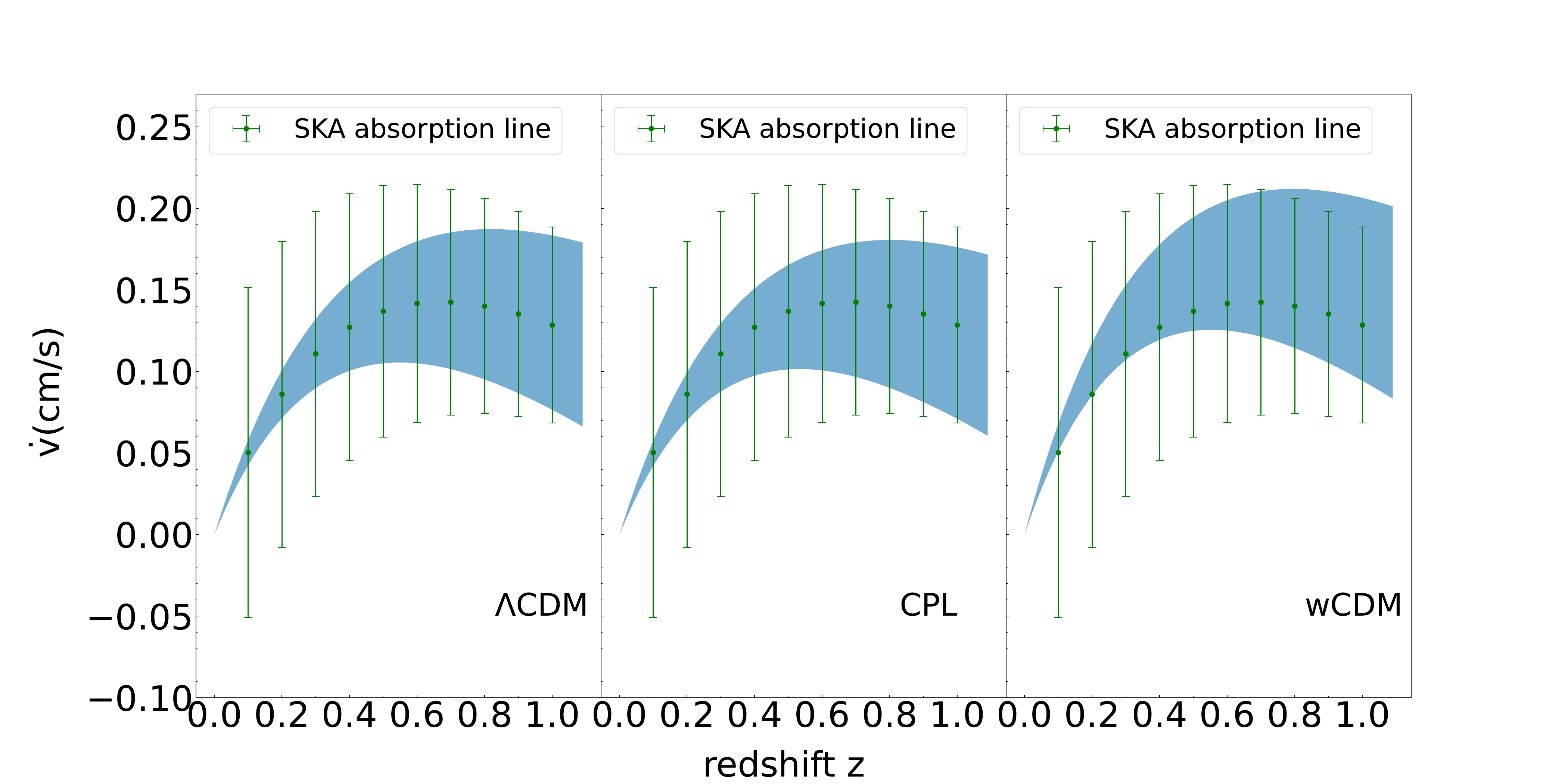}
	\caption{Theoretical velocity variations derived from three distinct models as $\rm H_0$ varies between 60 and 80 km/s/Mpc, depicted within the shallow blue regions. The error bars represent the empirical measurements of redshift drift and velocity drift obtained using the SKA via 21cm absorption lines over a 0.5-year observational period, within a redshift interval of 0 to 1.}\label{figs:obs4}
	
\end{figure}

\begin{figure}
	\includegraphics[width=.5\textwidth]{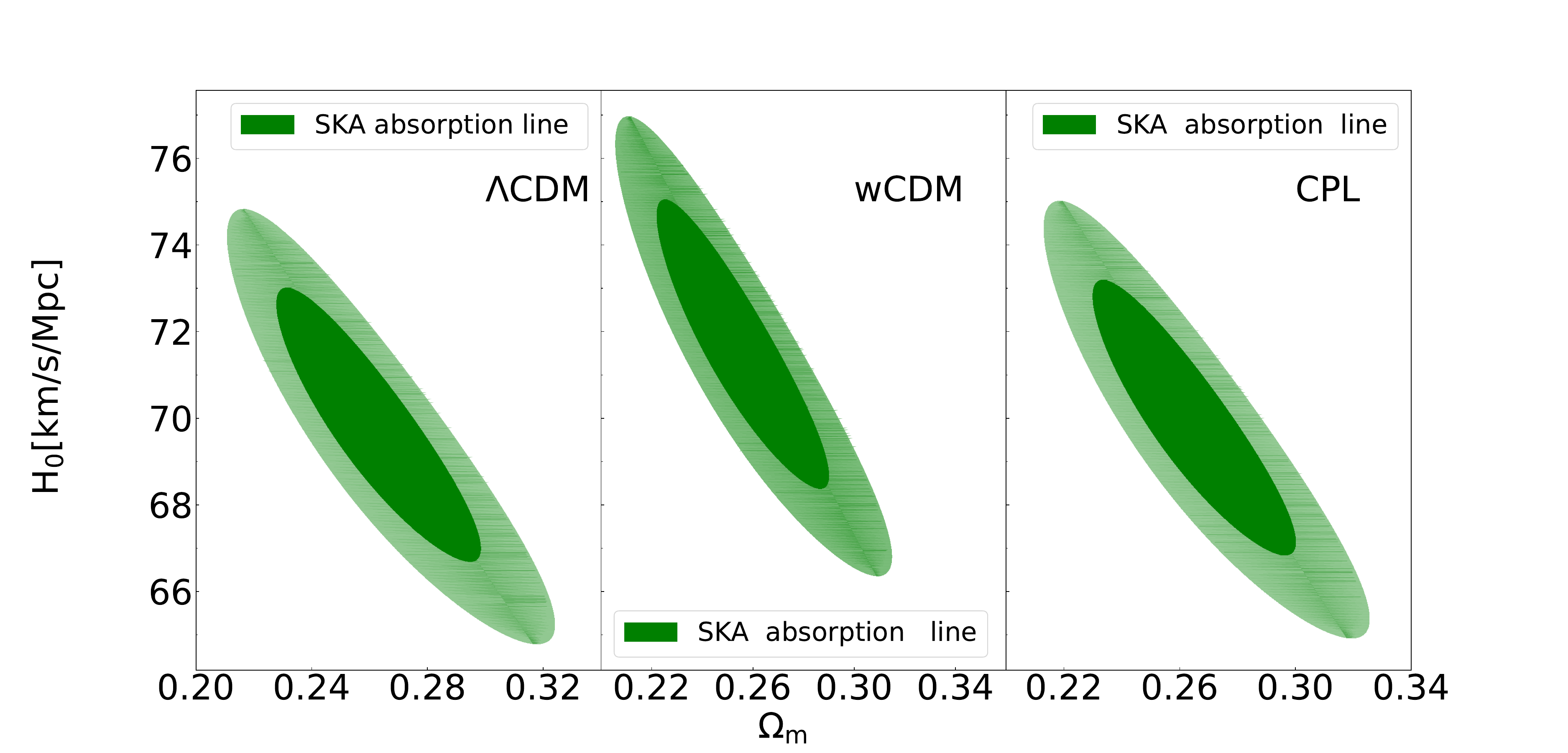}
	\caption{The best-fit value and 1$\sigma$(68.3\%) and 2$\sigma$(95.4\%) uncertainties of $H_0$ - $\Omega_m$ plane from SKA absorption line data under the three cosmological models.\label{figs:obs5}}
	
\end{figure}

\begin{figure}
	\includegraphics[width=.5\textwidth]{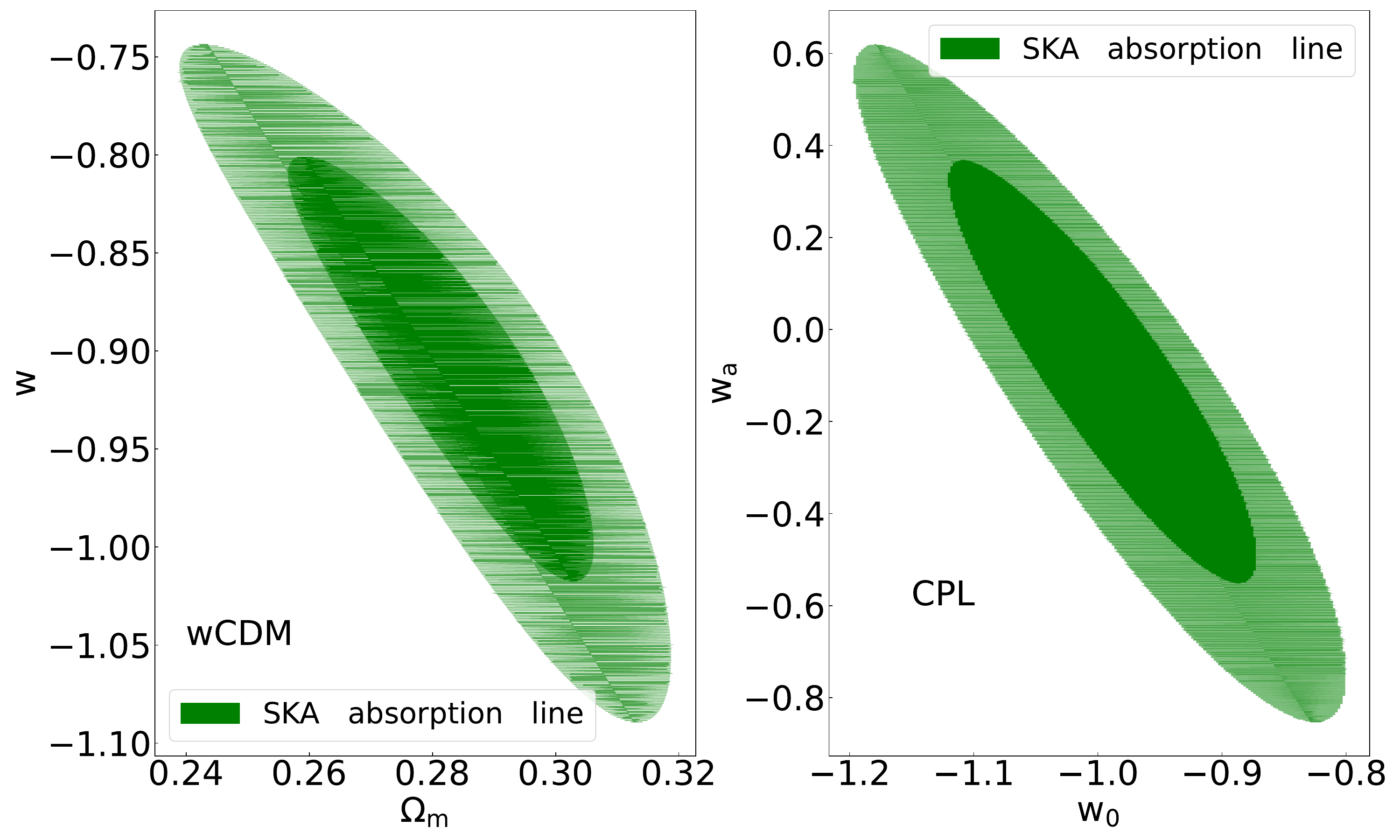}
	\caption{The best-fit value and 1$\sigma$(68.3\%) and 2$\sigma$(95.4\%) uncertainties of $\rm w$ and $\rm w_0 -w_a$ plane  from SKA absorption line data under the three cosmological models\label{figs:obs6}}.
	
\end{figure}

\begin{table*}
	 \caption{The best-fit value and 1$\sigma$(68.3\%) uncertainties  of $H_0$ and $\Omega_m$ from HI 21cm absorption line data of SKA  in the three cosmological models.\label{tabs:ab}}
	\centering
	\resizebox{\textwidth}{!}{
		\begin{tabular}{cccccc}
			\hline
			\hline
			
			model& $ H_0$ & $\Omega_m$ &w &$\rm w_0$ &$\rm w_a$ \\                     
			\hline
			$\Lambda$CDM&  $70.71^{+4.881}_{-3.046}$ &$0.274^{+0.057}_{-0.048}$ & -- & --  &--\\
			\hline         
			
			wCDM &$72.59^{+4.102}_{-5.397}$  & $0.284^{+0.146}_{-0.261}$  &   $-0.916^{+0.12}_{-0.107}$ & --  & --\\
			
			\hline

			CPL &$69.57^{+2.822}_{-4.453}$ & $0.293^{+0.271}_{-0.195}$ & --& $-0.974^{+0.141}_{-0.115}$ &$-0.142^{+0.372}_{-0.405}$\\
			\hline
			
	\end{tabular}}
\end{table*}

With respect to the analysis predicated on HI 21cm absorption, Figure \ref{figs:ab} delineates the enumeration of Damped Lyman Alpha (DLA) systems at an approximate redshift of z $\sim$ 1, as identified by the Square Kilometre Array (SKA) with a detection threshold sensitivity of 100mJy via the HI 21cm absorption line. The spatial density of DLA systems across a redshift interval from 0 to 5 is quantified in a specific functional form, particularly for those exhibiting an HI gas column density of at least $\rm 2 \times 10^{20}/cm^2$\citep{2004NewAR..48.1259K,2023MNRAS.521.3150L,2017MNRAS.471.3428R}:
 \begin{equation}\label{eqs:co}
 	\rm	n(z) = dN/dz =(0.027 ± 0.007)(1 + z)^{(1.682 \pm 0.2)}
 \end{equation}
 The prevalence of Damped Lyman Alpha (DLA) systems exhibits an escalation from 97.74 to 276.97 within the redshift interval of 0.1 to 1, ultimately culminating in a total count of 1818.75. The number density of DLAs is markedly diminished for redshifts less than 1.65. It is well-documented that the majority of DLA systems function as substantial repositories of neutral hydrogen in earlier cosmological periods, serving as progenitors to contemporary spiral galaxies. Additionally, the neutral hydrogen mass density contained within DLA systems at a redshift of approximately z $\sim$ 3 surpasses the presently observed levels by a factor of roughly four, paralleling the current stellar mass density observed in star-forming galaxies \citep{2004NewAR..48.1259K,2017MNRAS.471.3428R}. 
 
 The quantification of Damped Lyman-Alpha (DLA) systems enables the determination of the velocity drift ($\rm\dot{v}$), with an associated uncertainty $\sigma_i$ that scales inversely with the square root of the number of systems, denoted as $\rm 1/\sqrt{N}$. Figure \ref{figs:obs4} presents the projected velocity drift highlighted by green error bars, juxtaposed with the theoretical predictions for the Hubble constant ($\rm H_0$) shown within the light blue contour map, spanning a magnitude range of 0.05 to 0.10 cm/s over an observational period of $\Delta t =0.5$ years. The measurement uncertainties, ranging from 0.05 to 0.20 cm/s, are considerably larger than those obtained from emission data. This disparity is primarily due to the substantially lower number of observed DLA systems compared to the more prevalent HI 21cm emission line galaxies.

 The findings from the constrained analysis are presented in Table \ref{tabs:ab} and Figures \ref{figs:obs5} to \ref{figs:obs6}. Compared to the earlier results derived from the emission scenario, where all calculated values of $\rm H_0$ were below 70 with associated uncertainties not exceeding 1.5, the uncertainties related to the model parameters exhibit a notable increase. The estimations of $\rm H_0$ now range from 69.57 to 72.59 km/s/Mpc, with a minimum uncertainty surpassing 2.8. While $\rm\Omega_m$ remains slightly below 0.3, its associated uncertainties have also become more significant. Additionally, the confidence intervals for the dark energy equation of state parameters $\rm w, w_0, w_a$ at 1$\sigma$ and 2$\sigma$ deviations have significantly widened compared to the emission scenario results. The observed discrepancy stems from a significant paucity of HI absorption line systems relative to the abundant HI galaxies identified within the sample. Furthermore, to ascertain the constituents and diverse conditions of the HI gas, it is imperative to analyze the absorption line profiles through the application of multiple Gaussian fitting functions. High-resolution spectral data is essential for an intricate characterization of the HI gas structures detected via SKA's interferometric observations.

\begin{figure}
	\includegraphics[width=.5\textwidth]{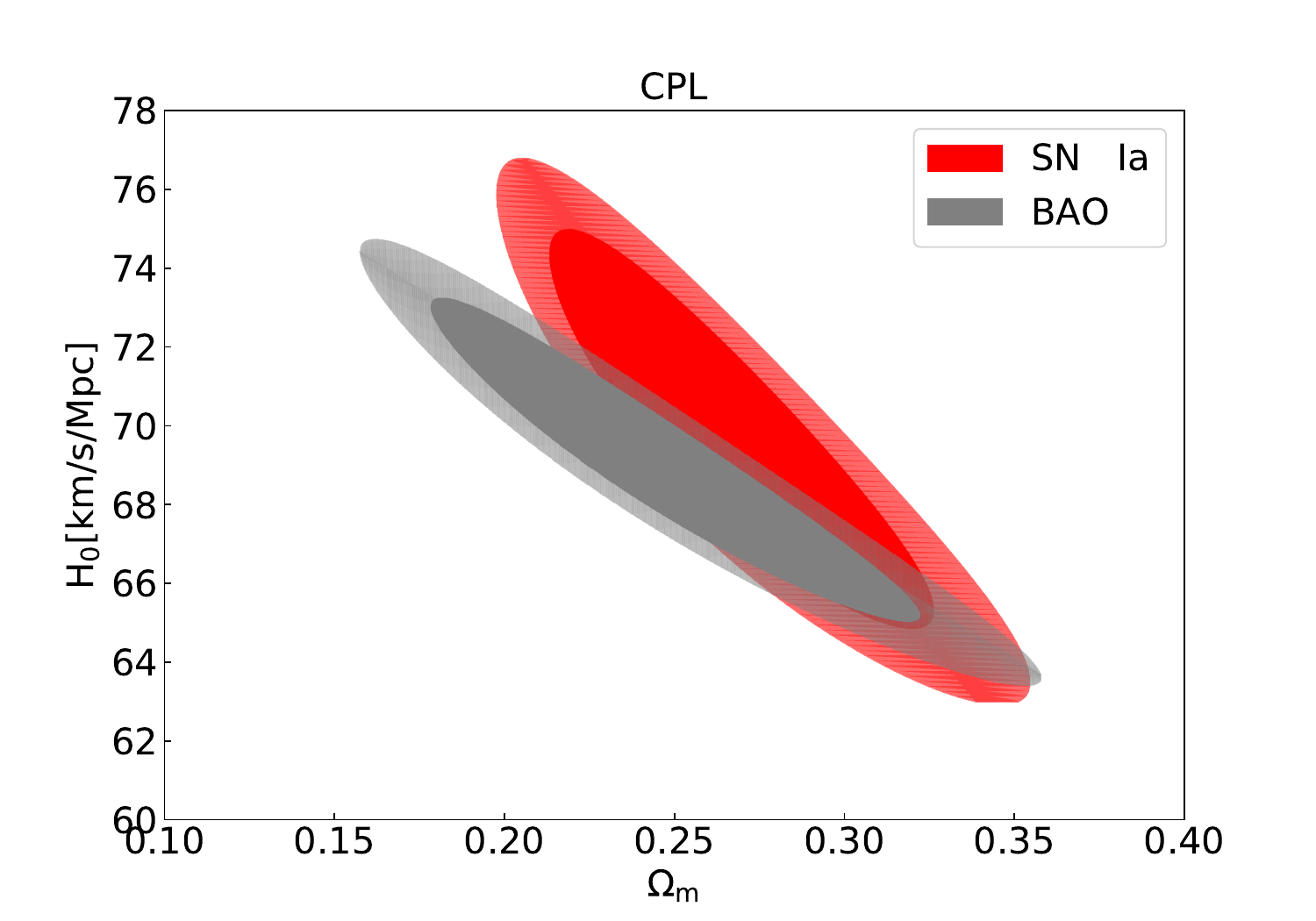}
	\caption{The 1$\sigma$(68.3\%) and 2$\sigma$(95.4\%) confidence level of $\rm H_0$ - $\rm \Omega_m$ plane constrainted results from SN Ia (red) and BAO data(gray)  under the  CPL model.\label{figs:com1}}
	
\end{figure}
 \begin{figure}
 	\includegraphics[width=.5\textwidth]{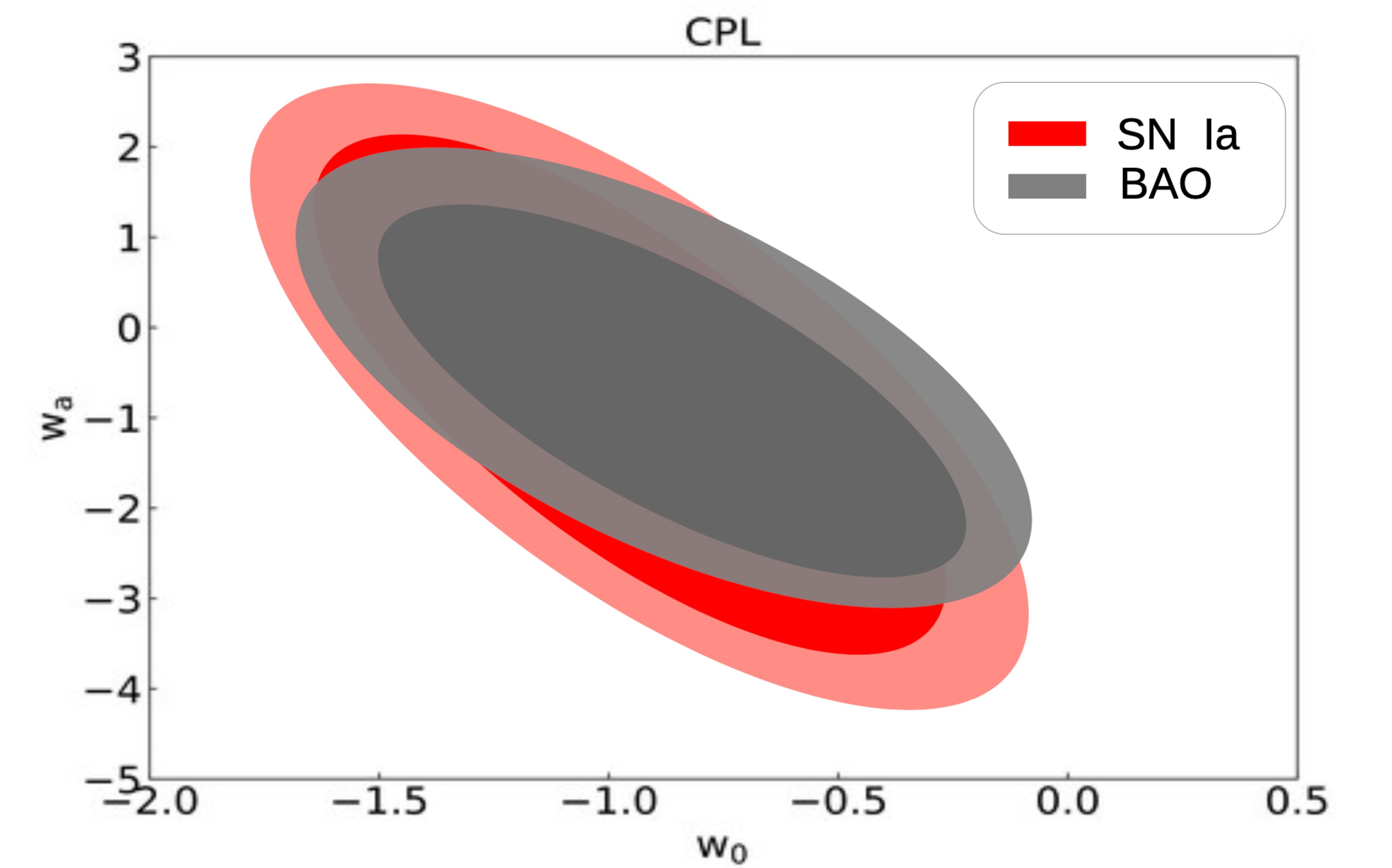}
 	\caption{The 1$\sigma$(68.3\%) and 2$\sigma$ confidence level of constraints on  $\rm w_0$-$\rm w_a$ plane using SN Ia (red) and BAO data(gray) under the  CPL model.\label{figs:com2}}
 	
 \end{figure}

To assess the efficacy of redshift drift in constraining cosmological models and parameters, akin to current leading probes such as Type Ia supernovae (SN Ia) and baryon acoustic oscillation (BAO) datasets, figures \ref{figs:com1} and \ref{figs:com2} illustrate the constrained cosmological parameters within the CPL framework. According to the SN Ia results \citep{2018ApJ...859..101S}, for the $H_0$ - $\Omega_m$ plane, the Hubble constant $H_0$ resides between 63 and 76.8 km/s/Mpc at a 2$\sigma$ confidence level (C.L.), while the matter density parameter $\Omega_m$ spans from 0.197 to 0.351. In the $\rm w_0$ - $\rm w_a$ parameter space, the constraints are somewhat more relaxed, with $\rm w_0$ ranging from -1.76 to -0.09 and $\rm w_a$ from -4.84 to 2.79. In comparison to the constrained results derived from 16 BAO data points as referenced in Table 1 of \citep{2020RAA....20...55K}, these findings exhibit slightly tighter constraints than those obtained from SN Ia, despite having fewer data points. Specifically, the $\Omega_m$ parameter spans a range from 0.15 to 0.352, which is marginally broader than the SN Ia fits, while the $H_0$ parameter varies between 64 and 74.6 km/s/Mpc. Additionally, $\rm w_0$ ranges from -1.5 to 0, and $\rm w_a$ ranges from -3.06 to 1.84. Consequently, integrating the conclusions from the two panels with the aforementioned redshift drift results, the cosmological implications become evident in terms of parameter constraints and precision estimation. This significantly narrows down the parameter ranges and elucidates the nature of dark energy, highlighting the advantages of these observations in cosmological studies.

\section{conclusion}\label{cons}
In this investigation, we initially determine the necessary observational accuracy for the cosmological redshift drift, denoted as $\dot{z}$, and the corresponding frequency drift, $\dot{\nu}$, within the context of the $\Lambda$CDM paradigm. This precision is assessed over observational time spans of $\Delta t$ equal to 0.5 and 1 year, as illustrated in figure \ref{figs:zh}, revealing that the redshift variations are approximately $\rm 10^{-11}$ and the frequency shift is below 0.1Hz for the same astronomical object, resulting in a velocity drift ranging from 0.5 to 3 mm/s per 0.5 year. Through a comparative analysis of the measured precision of redshift drift against the HI 21cm signal detections by the SKA at four distinct high spectral resolutions,namely, 0.001, 0.002, 0.005, and 0.01 Hz, it was ascertained that resolutions at 0.001 Hz and 0.002 Hz suffice to achieve the requisite precision for dimensionless redshift and velocity drift, as demonstrated in Figures \ref{figs:sz} and \ref{figs:sv} and Table \ref{tabs:sensi}. Furthermore, the redshift interval of galaxies detected by the SKA via HI 21cm line emission in this research can be delineated. Specifically, face-on galaxies exhibiting a distinctive Gaussian flux profile are likely candidates for this signal. The magnitude of velocity drift, along with its associated uncertainties calculated based on the inverse square root of the number of sources ($\rm 1/\sqrt{N}$), is illustrated in Figure \ref{figs:num}. The observed velocity drift ranges from 0.05 to 0.15 cm/s per half year, with corresponding uncertainties of 2 to 5 mm/s over a redshift span from 0 to 1. As a result, we utilize these outcomes to enhance the prevailing cosmological models and parameters. The inferred value of $H_0$ is approximately 70 km/s/Mpc, $\Omega_m$ remains consistently around 0.3, $w$ is near -1, $w_0$ is close to -1, and $w_a$ approaches -0.1, regardless of the various models employed. These findings are in strict concordance with previous data. Conversely, the Damped Lyman Alpha (DLA) systems, consisting of cold, dense HI gas clouds detectable by the Square Kilometre Array (SKA), are not constrained by redshift limitations and exhibit a distinct optical depth signature in the spectra of background quasars or radio galaxies, manifesting as redshifted 21cm absorption lines. Figure \ref{figs:ab} depicts the distribution of DLA systems, along with the precision indicated by the error bars in Figure \ref{figs:obs4}.

The aggregate total of identified DLA systems has reached 1800, which permits the imposition of stringent constraints on cosmological parameters, as illustrated in figures \ref{figs:obs5} and \ref{figs:obs6}. In this context, the predicted velocity drift range, characterized by fluctuations from 0.05 to 0.20 cm/s per 0.5 year, exhibits marginal uncertainties that are comparatively broader than those inferred from emission data. Nonetheless, the conclusive evidence underscores the singularity of the redshift drift phenomenon as a critical diagnostic tool for advancing cosmological research in the SKA epoch, and for deepening our understanding of the fundamental characteristics of dark energy. It is anticipated that the redshift drift signal can be accurately measured at the precision of millimeters per second using spectral resolution parameters of 0.001Hz and 0.002Hz over an observation duration of $\Delta t$ = 0.5 years. The elucidation of cosmic acceleration driven by dark energy can ultimately be realized through either the analysis of HI 21cm emissions from galaxies or the examination of the absorption line profiles of Damped Lyman Alpha (DLA) systems.

\section{Acknowledgements}
This work was supported by National SKA Program of China (2022SKA0110202) and China Manned Space Program through its Space Application System.

\bibliography{ska}

\begin{thebibliography}{}
\expandafter\ifx\csname natexlab\endcsname\relax\def\natexlab#1{#1}\fi
\providecommand{\url}[1]{\href{#1}{#1}}
\providecommand{\dodoi}[1]{doi:~\href{http://doi.org/#1}{\nolinkurl{#1}}}
\providecommand{\doeprint}[1]{\href{http://ascl.net/#1}{\nolinkurl{http://ascl.net/#1}}}
\providecommand{\doarXiv}[1]{\href{https://arxiv.org/abs/#1}{\nolinkurl{https://arxiv.org/abs/#1}}}

\bibitem[{{Abdalla} {et~al.}(2015){Abdalla}, {Bull}, {Camera},
  {Benoit-L{\'e}vy}, {Joachimi}, {Kirk}, {Kloeckner}, {Maartens}, {Raccanelli},
  {Santos}, \& {Zhao}}]{2015aska.confE..17A}
{Abdalla}, F.~B., {Bull}, P., {Camera}, S., {et~al.} 2015, in Advancing
  Astrophysics with the Square Kilometre Array (AASKA14), 17,
  \dodoi{10.22323/1.215.0017}

\bibitem[{{Allison} {et~al.}(2022){Allison}, {Sadler}, {Amaral}, {An},
  {Curran}, {Darling}, {Edge}, {Ellison}, {Emig}, {Gaensler},
  {Garratt-Smithson}, {Glowacki}, {Grasha}, {Koribalski}, {Lagos}, {Lah},
  {Mahony}, {Mao}, {Morganti}, {Moss}, {Pettini}, {Pimbblet}, {Power}, {Salas},
  {Staveley-Smith}, {Whiting}, {Wong}, {Yoon}, {Zheng}, \&
  {Zwaan}}]{2022PASA...39...10A}
{Allison}, J.~R., {Sadler}, E.~M., {Amaral}, A.~D., {et~al.} 2022, \pasa, 39,
  e010, \dodoi{10.1017/pasa.2022.3}

\bibitem[{{Alves} {et~al.}(2019){Alves}, {Leite}, {Martins}, {Matos}, \&
  {Silva}}]{2019MNRAS.488.3607A}
{Alves}, C.~S., {Leite}, A.~C.~O., {Martins}, C.~J.~A.~P., {Matos}, J.~G.~B.,
  \& {Silva}, T.~A. 2019, \mnras, 488, 3607, \dodoi{10.1093/mnras/stz1934}

\bibitem[{Alves {et~al.}(2019)Alves, Leite, Martins, Matos, \& Silva}]{Alves}
Alves, C.~S., Leite, A. C.~O., Martins, C. J. A.~P., Matos, J. G.~B., \& Silva,
  T.~A. 2019, Mon. Not. Roy. Astron. Soc., 488, 3607,
  \dodoi{10.1093/mnras/stz1934}

\bibitem[{{Balbi} \& {Quercellini}(2007)}]{2007MNRAS.382.1623B}
{Balbi}, A., \& {Quercellini}, C. 2007, \mnras, 382, 1623,
  \dodoi{10.1111/j.1365-2966.2007.12407.x}

\bibitem[{{Bolejko} {et~al.}(2019){Bolejko}, {Wang}, \&
  {Lewis}}]{2019arXiv190704495B}
{Bolejko}, K., {Wang}, C., \& {Lewis}, G.~F. 2019, arXiv e-prints,
  arXiv:1907.04495, \dodoi{10.48550/arXiv.1907.04495}

\bibitem[{{Braun} {et~al.}(2019){Braun}, {Bonaldi}, {Bourke}, {Keane}, \&
  {Wagg}}]{2019arXiv191212699B}
{Braun}, R., {Bonaldi}, A., {Bourke}, T., {Keane}, E., \& {Wagg}, J. 2019,
  arXiv e-prints, arXiv:1912.12699, \dodoi{10.48550/arXiv.1912.12699}

\bibitem[{Caldwell \&
  Kamionkowski(2009)}]{annurev:/content/journals/10.1146/annurev-nucl-010709-151330}
Caldwell, R.~R., \& Kamionkowski, M. 2009, Annual Review of Nuclear and
  Particle Science, 59, 397,
  \dodoi{https://doi.org/10.1146/annurev-nucl-010709-151330}

\bibitem[{Cooke(2019)}]{Cooke_2019}
Cooke, R. 2019, Monthly Notices of the Royal Astronomical Society, 492, 2044,
  \dodoi{10.1093/mnras/stz3465}

\bibitem[{{Cooke}(2020)}]{2020MNRAS.492.2044C}
{Cooke}, R. 2020, \mnras, 492, 2044, \dodoi{10.1093/mnras/stz3465}

\bibitem[{{Cotsakis} {et~al.}(2023){Cotsakis}, {Mimoso}, \&
  {Miritzis}}]{2023EPJC...83..735C}
{Cotsakis}, S., {Mimoso}, J.~P., \& {Miritzis}, J. 2023, European Physical
  Journal C, 83, 735, \dodoi{10.1140/epjc/s10052-023-11922-z}

\bibitem[{{Cristiani} {et~al.}(2023){Cristiani}, {Boutsia}, {Calderone},
  {Cupani}, {D'Odorico}, {Fontanot}, {Grazian}, {Guarneri}, {Martins},
  {Pasquini}, {Porru}, \& {Vanzella}}]{2023arXiv230204365C}
{Cristiani}, S., {Boutsia}, K., {Calderone}, G., {et~al.} 2023, arXiv e-prints,
  arXiv:2302.04365, \dodoi{10.48550/arXiv.2302.04365}

\bibitem[{Darling(2012)}]{Darling_2012}
Darling, J. 2012, The Astrophysical Journal, 761, L26,
  \dodoi{10.1088/2041-8205/761/2/l26}

\bibitem[{{Dong} {et~al.}(2022){Dong}, {Gonzalez}, {Eikenberry}, {Jeram},
  {Likamonsavad}, {Liske}, {Stelter}, \& {Townsend}}]{2022MNRAS.514.5493D}
{Dong}, C., {Gonzalez}, A., {Eikenberry}, S., {et~al.} 2022, \mnras, 514, 5493,
  \dodoi{10.1093/mnras/stac1702}

\bibitem[{{Dutta} {et~al.}(2022){Dutta}, {Kurapati}, {Aditya}, {Bait}, {Das},
  {Dutta}, {Indulekha}, {Nandakumar}, {Patra}, {Roy}, \&
  {Roychowdhury}}]{2022JApA...43..103D}
{Dutta}, R., {Kurapati}, S., {Aditya}, J.~N.~H.~S., {et~al.} 2022, Journal of
  Astrophysics and Astronomy, 43, 103, \dodoi{10.1007/s12036-022-09875-y}

\bibitem[{{Eden} {et~al.}(2024){Eden}, {Sadler}, {Pimbblet}, {Mahony}, \&
  {Yoon}}]{2024MNRAS.tmp.2497E}
{Eden}, S.~L., {Sadler}, E.~M., {Pimbblet}, K.~A., {Mahony}, E.~K., \& {Yoon},
  H. 2024, \mnras, \dodoi{10.1093/mnras/stae2581}

\bibitem[{{Esteves} {et~al.}(2021){Esteves}, {Martins}, {Pereira}, \&
  {Alves}}]{2021MNRAS.508L..53E}
{Esteves}, J., {Martins}, C.~J.~A.~P., {Pereira}, B.~G., \& {Alves}, C.~S.
  2021, \mnras, 508, L53, \dodoi{10.1093/mnrasl/slab102}

\bibitem[{{Ger{\'e}b} {et~al.}(2015){Ger{\'e}b}, {Maccagni}, {Morganti}, \&
  {Oosterloo}}]{2015A&A...575A..44G}
{Ger{\'e}b}, K., {Maccagni}, F.~M., {Morganti}, R., \& {Oosterloo}, T.~A. 2015,
  \aap, 575, A44, \dodoi{10.1051/0004-6361/201424655}

\bibitem[{{Guo} \& {Zhang}(2016)}]{2016EPJC...76..163G}
{Guo}, R.-Y., \& {Zhang}, X. 2016, European Physical Journal C, 76, 163,
  \dodoi{10.1140/epjc/s10052-016-4016-x}

\bibitem[{Gupta {et~al.}(2013)Gupta, Srianand, Noterdaeme, Petitjean, \&
  Muzahid}]{Gupta_2013}
Gupta, N., Srianand, R., Noterdaeme, P., Petitjean, P., \& Muzahid, S. 2013,
  \aap, 558, A84, \dodoi{10.1051/0004-6361/201321609}

\bibitem[{{Heinesen}(2021)}]{2021PhRvD.103h1302H}
{Heinesen}, A. 2021, \prd, 103, L081302, \dodoi{10.1103/PhysRevD.103.L081302}

\bibitem[{Jiao {et~al.}(2020)Jiao, Zhang, Zhang, Yu, Zhu, \& Li}]{Jiao_2020}
Jiao, K., Zhang, J.-C., Zhang, T.-J., {et~al.} 2020, Journal of Cosmology and
  Astroparticle Physics, 2020, 054, \dodoi{10.1088/1475-7516/2020/01/054}

\bibitem[{{Kanekar} \& {Briggs}(2004)}]{2004NewAR..48.1259K}
{Kanekar}, N., \& {Briggs}, F.~H. 2004, \nar, 48, 1259,
  \dodoi{10.1016/j.newar.2004.09.030}

\bibitem[{{Kanekar} \& {Chengalur}(2001)}]{2001A&A...369...42K}
{Kanekar}, N., \& {Chengalur}, J.~N. 2001, \aap, 369, 42,
  \dodoi{10.1051/0004-6361:20010096}

\bibitem[{Kanekar {et~al.}(2001)Kanekar, Ghosh, \& Chengalur}]{Kanekar_2001}
Kanekar, N., Ghosh, T., \& Chengalur, J.~N. 2001, Astronomy {\&} Astrophysics,
  373, 394, \dodoi{10.1051/0004-6361:20010545}

\bibitem[{{Kanekar} {et~al.}(2001){Kanekar}, {Ghosh}, \&
  {Chengalur}}]{2001A&A...373..394K}
{Kanekar}, N., {Ghosh}, T., \& {Chengalur}, J.~N. 2001, \aap, 373, 394,
  \dodoi{10.1051/0004-6361:20010545}

\bibitem[{{Kang}(2021)}]{2021PDU....3100784K}
{Kang}, J. 2021, Physics of the Dark Universe, 31, 100784,
  \dodoi{10.1016/j.dark.2021.100784}

\bibitem[{{Kang} {et~al.}(2023){Kang}, {Lu}, {Zhang}, \&
  {Zhu}}]{2023arXiv230808851K}
{Kang}, J., {Lu}, C.-Z., {Zhang}, T., \& {Zhu}, M. 2023, arXiv e-prints,
  arXiv:2308.08851, \dodoi{10.48550/arXiv.2308.08851}

\bibitem[{{Kang} {et~al.}(2024){Kang}, {Lu}, {Zhang}, \&
  {Zhu}}]{2024RAA....24g5002K}
{Kang}, J., {Lu}, C.-Z., {Zhang}, T.-J., \& {Zhu}, M. 2024, Research in
  Astronomy and Astrophysics, 24, 075002, \dodoi{10.1088/1674-4527/ad48d1}

\bibitem[{{Kang} {et~al.}(2020){Kang}, {Gong}, {Cheng}, \&
  {Chen}}]{2020RAA....20...55K}
{Kang}, J.-G., {Gong}, Y., {Cheng}, G., \& {Chen}, X. 2020, Research in
  Astronomy and Astrophysics, 20, 055, \dodoi{10.1088/1674-4527/20/4/55}

\bibitem[{{Kim} {et~al.}(2015){Kim}, {Linder}, {Edelstein}, \&
  {Erskine}}]{2015APh....62..195K}
{Kim}, A.~G., {Linder}, E.~V., {Edelstein}, J., \& {Erskine}, D. 2015,
  Astroparticle Physics, 62, 195, \dodoi{10.1016/j.astropartphys.2014.09.004}

\bibitem[{{Kloeckner} {et~al.}(2015){Kloeckner}, {Obreschkow}, {Martins},
  {Raccanelli}, {Champion}, {Roy}, {Lobanov}, {Wagner}, \&
  {Keller}}]{2015aska.confE..27K}
{Kloeckner}, H.~R., {Obreschkow}, D., {Martins}, C., {et~al.} 2015, in
  Advancing Astrophysics with the Square Kilometre Array (AASKA14), 27,
  \dodoi{10.22323/1.215.0027}

\bibitem[{{Lane} {et~al.}(1998){Lane}, {Smette}, {Briggs}, {Rao}, {Turnshek},
  \& {Meylan}}]{1998AJ....116...26L}
{Lane}, W., {Smette}, A., {Briggs}, F., {et~al.} 1998, \aj, 116, 26,
  \dodoi{10.1086/300422}

\bibitem[{Lazio(2009)}]{lazio2009square}
Lazio, J. 2009, The Square Kilometre Array.
\newblock \doarXiv{0910.0632}

\bibitem[{{Liske} {et~al.}(2008){Liske}, {Grazian}, {Vanzella}, {Dessauges},
  {Viel}, {Pasquini}, {Haehnelt}, {Cristiani}, {Pepe}, {Avila}, {Bonifacio},
  {Bouchy}, {Dekker}, {Delabre}, {D'Odorico}, {D'Odorico}, {Levshakov},
  {Lovis}, {Mayor}, {Molaro}, {Moscardini}, {Murphy}, {Queloz}, {Shaver},
  {Udry}, {Wiklind}, \& {Zucker}}]{2008MNRAS.386.1192L}
{Liske}, J., {Grazian}, A., {Vanzella}, E., {et~al.} 2008, \mnras, 386, 1192,
  \dodoi{10.1111/j.1365-2966.2008.13090.x}

\bibitem[{{Liu} {et~al.}(2020){Liu}, {Zhang}, \& {Zhang}}]{2020EPJC...80..304L}
{Liu}, Y., {Zhang}, J.-F., \& {Zhang}, X. 2020, European Physical Journal C,
  80, 304, \dodoi{10.1140/epjc/s10052-020-7863-4}

\bibitem[{{Loeb}(1998)}]{1998ApJ...499L.111L}
{Loeb}, A. 1998, \apjl, 499, L111, \dodoi{10.1086/311375}

\bibitem[{{Lu} {et~al.}(2022){Lu}, {Jiao}, {Zhang}, {Zhang}, \&
  {Zhu}}]{2022PDU....3701088L}
{Lu}, C.-Z., {Jiao}, K., {Zhang}, T., {Zhang}, T.-J., \& {Zhu}, M. 2022,
  Physics of the Dark Universe, 37, 101088, \dodoi{10.1016/j.dark.2022.101088}

\bibitem[{{Lu} {et~al.}(2023){Lu}, {Zhang}, \& {Zhang}}]{2023MNRAS.521.3150L}
{Lu}, C.-Z., {Zhang}, T., \& {Zhang}, T.-J. 2023, \mnras, 521, 3150,
  \dodoi{10.1093/mnras/stad761}

\bibitem[{{Martinelli} {et~al.}(2012){Martinelli}, {Pandolfi}, {Martins}, \&
  {Vielzeuf}}]{2012PhRvD..86l3001M}
{Martinelli}, M., {Pandolfi}, S., {Martins}, C.~J.~A.~P., \& {Vielzeuf}, P.~E.
  2012, \prd, 86, 123001, \dodoi{10.1103/PhysRevD.86.123001}

\bibitem[{{Martins} {et~al.}(2021){Martins}, {Alves}, {Esteves}, {Lapel}, \&
  {Pereira}}]{2021arXiv211012242M}
{Martins}, C.~J.~A.~P., {Alves}, C.~S., {Esteves}, J., {Lapel}, A., \&
  {Pereira}, B.~G. 2021, arXiv e-prints, arXiv:2110.12242.
\newblock \doarXiv{2110.12242}

\bibitem[{{Mishra} \& {C{\'e}l{\'e}rier}(2015)}]{2015mgm..conf.1590M}
{Mishra}, P., \& {C{\'e}l{\'e}rier}, Marie-No{\"e}lle~Singh, T.~P. 2015, in
  Thirteenth Marcel Grossmann Meeting: On Recent Developments in Theoretical
  and Experimental General Relativity, Astrophysics and Relativistic Field
  Theories, 1590--1592, \dodoi{10.1142/9789814623995_0233}

\bibitem[{Moresco {et~al.}(2022)Moresco, Amati, Amendola, Birrer, Blakeslee,
  Cantiello, Cimatti, Darling, Della~Valle, Fishbach, Grillo, Hamaus, Holz,
  Izzo, Jimenez, Lusso, Meneghetti, Piedipalumbo, Pisani, Pourtsidou, Pozzetti,
  Quartin, Risaliti, Rosati, \& Verde}]{Moresco_2022}
Moresco, M., Amati, L., Amendola, L., {et~al.} 2022, Living Reviews in
  Relativity, 25, \dodoi{10.1007/s41114-022-00040-z}

\bibitem[{{Morganti} {et~al.}(2015){Morganti}, {Sadler}, \&
  {Curran}}]{2015aska.confE.134M}
{Morganti}, R., {Sadler}, E.~M., \& {Curran}, S. 2015, in Advancing
  Astrophysics with the Square Kilometre Array (AASKA14), 134,
  \dodoi{10.22323/1.215.0134}

\bibitem[{{Obreschkow} {et~al.}(2009){Obreschkow}, {Kl{\"o}ckner}, {Heywood},
  {Levrier}, \& {Rawlings}}]{2009ApJ...703.1890O}
{Obreschkow}, D., {Kl{\"o}ckner}, H.~R., {Heywood}, I., {Levrier}, F., \&
  {Rawlings}, S. 2009, \apj, 703, 1890, \dodoi{10.1088/0004-637X/703/2/1890}

\bibitem[{{Perlmutter} {et~al.}(1999){Perlmutter}, {Turner}, \&
  {White}}]{1999PhRvL..83..670P}
{Perlmutter}, S., {Turner}, M.~S., \& {White}, M. 1999, \prl, 83, 670,
  \dodoi{10.1103/PhysRevLett.83.670}

\bibitem[{{Planck Collaboration} {et~al.}(2020){Planck Collaboration},
  {Aghanim}, {Akrami}, {Ashdown}, {Aumont}, {Baccigalupi}, {Ballardini},
  {Banday}, {Barreiro}, {Bartolo}, {Basak}, {Battye}, {Benabed}, {Bernard},
  {Bersanelli}, {Bielewicz}, {Bock}, {Bond}, {Borrill}, {Bouchet}, {Boulanger},
  {Bucher}, {Burigana}, {Butler}, {Calabrese}, {Cardoso}, {Carron},
  {Challinor}, {Chiang}, {Chluba}, {Colombo}, {Combet}, {Contreras}, {Crill},
  {Cuttaia}, {de Bernardis}, {de Zotti}, {Delabrouille}, {Delouis}, {Di
  Valentino}, {Diego}, {Dor{\'e}}, {Douspis}, {Ducout}, {Dupac}, {Dusini},
  {Efstathiou}, {Elsner}, {En{\ss}lin}, {Eriksen}, {Fantaye}, {Farhang},
  {Fergusson}, {Fernandez-Cobos}, {Finelli}, {Forastieri}, {Frailis},
  {Fraisse}, {Franceschi}, {Frolov}, {Galeotta}, {Galli}, {Ganga},
  {G{\'e}nova-Santos}, {Gerbino}, {Ghosh}, {Gonz{\'a}lez-Nuevo}, {G{\'o}rski},
  {Gratton}, {Gruppuso}, {Gudmundsson}, {Hamann}, {Handley}, {Hansen},
  {Herranz}, {Hildebrandt}, {Hivon}, {Huang}, {Jaffe}, {Jones}, {Karakci},
  {Keih{\"a}nen}, {Keskitalo}, {Kiiveri}, {Kim}, {Kisner}, {Knox},
  {Krachmalnicoff}, {Kunz}, {Kurki-Suonio}, {Lagache}, {Lamarre}, {Lasenby},
  {Lattanzi}, {Lawrence}, {Le Jeune}, {Lemos}, {Lesgourgues}, {Levrier},
  {Lewis}, {Liguori}, {Lilje}, {Lilley}, {Lindholm}, {L{\'o}pez-Caniego},
  {Lubin}, {Ma}, {Mac{\'\i}as-P{\'e}rez}, {Maggio}, {Maino}, {Mandolesi},
  {Mangilli}, {Marcos-Caballero}, {Maris}, {Martin}, {Martinelli},
  {Mart{\'\i}nez-Gonz{\'a}lez}, {Matarrese}, {Mauri}, {McEwen}, {Meinhold},
  {Melchiorri}, {Mennella}, {Migliaccio}, {Millea}, {Mitra},
  {Miville-Desch{\^e}nes}, {Molinari}, {Montier}, {Morgante}, {Moss}, {Natoli},
  {N{\o}rgaard-Nielsen}, {Pagano}, {Paoletti}, {Partridge}, {Patanchon},
  {Peiris}, {Perrotta}, {Pettorino}, {Piacentini}, {Polastri}, {Polenta},
  {Puget}, {Rachen}, {Reinecke}, {Remazeilles}, {Renzi}, {Rocha}, {Rosset},
  {Roudier}, {Rubi{\~n}o-Mart{\'\i}n}, {Ruiz-Granados}, {Salvati}, {Sandri},
  {Savelainen}, {Scott}, {Shellard}, {Sirignano}, {Sirri}, {Spencer},
  {Sunyaev}, {Suur-Uski}, {Tauber}, {Tavagnacco}, {Tenti}, {Toffolatti},
  {Tomasi}, {Trombetti}, {Valenziano}, {Valiviita}, {Van Tent}, {Vibert},
  {Vielva}, {Villa}, {Vittorio}, {Wandelt}, {Wehus}, {White}, {White},
  {Zacchei}, \& {Zonca}}]{2020A&A...641A...6P}
{Planck Collaboration}, {Aghanim}, N., {Akrami}, Y., {et~al.} 2020, \aap, 641,
  A6, \dodoi{10.1051/0004-6361/201833910}

\bibitem[{Pritchard \& Loeb(2012)}]{Pritchard_2012}
Pritchard, J.~R., \& Loeb, A. 2012, Reports on Progress in Physics, 75, 086901,
  \dodoi{10.1088/0034-4885/75/8/086901}

\bibitem[{Quercellini {et~al.}(2012)Quercellini, Amendola, Balbi, Cabella, \&
  Quartin}]{Quercellini_2012}
Quercellini, C., Amendola, L., Balbi, A., Cabella, P., \& Quartin, M. 2012,
  Physics Reports, 521, 95, \dodoi{10.1016/j.physrep.2012.09.002}

\bibitem[{{Rao} {et~al.}(2017){Rao}, {Turnshek}, {Sardane}, \&
  {Monier}}]{2017MNRAS.471.3428R}
{Rao}, S.~M., {Turnshek}, D.~A., {Sardane}, G.~M., \& {Monier}, E.~M. 2017,
  \mnras, 471, 3428, \dodoi{10.1093/mnras/stx1787}

\bibitem[{Rawlings \& Schilizzi(2011)}]{rawlings2011square}
Rawlings, S., \& Schilizzi, R. 2011, The Square Kilometre Array.
\newblock \doarXiv{1105.5953}

\bibitem[{{Rawlings} \& {Schilizzi}(2011)}]{2011arXiv1105.5953R}
{Rawlings}, S., \& {Schilizzi}, R. 2011, arXiv e-prints, arXiv:1105.5953,
  \dodoi{10.48550/arXiv.1105.5953}

\bibitem[{{Riess} {et~al.}(1998){Riess}, {Filippenko}, {Challis},
  {Clocchiatti}, {Diercks}, {Garnavich}, {Gilliland}, {Hogan}, {Jha},
  {Kirshner}, {Leibundgut}, {Phillips}, {Reiss}, {Schmidt}, {Schommer},
  {Smith}, {Spyromilio}, {Stubbs}, {Suntzeff}, \&
  {Tonry}}]{1998AJ....116.1009R}
{Riess}, A.~G., {Filippenko}, A.~V., {Challis}, P., {et~al.} 1998, \aj, 116,
  1009, \dodoi{10.1086/300499}

\bibitem[{{Rocha} \& {Martins}(2023)}]{2023MNRAS.518.2853R}
{Rocha}, B.~A.~R., \& {Martins}, C.~J.~A.~P. 2023, \mnras, 518, 2853,
  \dodoi{10.1093/mnras/stac3240}

\bibitem[{{Sandage}(1962)}]{1962ApJ...136..319S}
{Sandage}, A. 1962, \apj, 136, 319, \dodoi{10.1086/147385}

\bibitem[{{Scolnic} {et~al.}(2018){Scolnic}, {Jones}, {Rest}, {Pan},
  {Chornock}, {Foley}, {Huber}, {Kessler}, {Narayan}, {Riess}, {Rodney},
  {Berger}, {Brout}, {Challis}, {Drout}, {Finkbeiner}, {Lunnan}, {Kirshner},
  {Sanders}, {Schlafly}, {Smartt}, {Stubbs}, {Tonry}, {Wood-Vasey}, {Foley},
  {Hand}, {Johnson}, {Burgett}, {Chambers}, {Draper}, {Hodapp}, {Kaiser},
  {Kudritzki}, {Magnier}, {Metcalfe}, {Bresolin}, {Gall}, {Kotak}, {McCrum}, \&
  {Smith}}]{2018ApJ...859..101S}
{Scolnic}, D.~M., {Jones}, D.~O., {Rest}, A., {et~al.} 2018, \apj, 859, 101,
  \dodoi{10.3847/1538-4357/aab9bb}

\bibitem[{{Spekkens} {et~al.}(2019){Spekkens}, {Chiang}, {Kothes},
  {Rosolowsky}, {Rupen}, {Safi-Harb}, {Sievers}, {Stairs}, {van der Marel},
  {Abraham}, {Alexandroff}, {Bartel}, {Baum}, {Bietenholz}, {Boley}, {Bond},
  {Brown}, {Brown}, {Davis}, {English}, {Fahlman}, {Ferrarese}, {Di Francesco},
  {Gaensler}, {Gaudet}, {Graber}, {Halpern}, {Hill}, {Hlavacek-Larrondo},
  {Irwin}, {Johnstone}, {Joncas}, {Kaspi}, {Kavelaars}, {Liu}, {Matthews},
  {Mirocha}, {Monsalve}, {Ng}, {O'Dea}, {Pen}, {Plume}, {Robishaw}, {Sadavoy},
  {Sanghai}, {Scholz}, {Simard}, {Shaw}, {Singh}, {Sigurdson}, {Stevens},
  {Stil}, {Tulin}, {van Eck}, {Wall}, {West}, {Woods}, \&
  {Wulf}}]{2019clrp.2020...46S}
{Spekkens}, K., {Chiang}, C., {Kothes}, R., {et~al.} 2019, in Canadian Long
  Range Plan for Astronomy and Astrophysics White Papers, Vol. 2020, 46,
  \dodoi{10.5281/zenodo.3825168}

\bibitem[{{Square Kilometre Array Cosmology Science Working Group}
  {et~al.}(2020){Square Kilometre Array Cosmology Science Working Group},
  {Bacon}, {Battye}, {Bull}, {Camera}, {Ferreira}, {Harrison}, {Parkinson},
  {Pourtsidou}, {Santos}, {Wolz}, {Abdalla}, {Akrami}, {Alonso},
  {Andrianomena}, {Ballardini}, {Bernal}, {Bertacca}, {Bengaly}, {Bonaldi},
  {Bonvin}, {Brown}, {Chapman}, {Chen}, {Chen}, {Cunnington}, {Davis},
  {Dickinson}, {Fonseca}, {Grainge}, {Harper}, {Jarvis}, {Maartens}, {Maddox},
  {Padmanabhan}, {Pritchard}, {Raccanelli}, {Rivi}, {Roychowdhury},
  {Sahl{\'e}n}, {Schwarz}, {Siewert}, {Viel}, {Villaescusa-Navarro}, {Xu},
  {Yamauchi}, \& {Zuntz}}]{2020PASA...37....7S}
{Square Kilometre Array Cosmology Science Working Group}, {Bacon}, D.~J.,
  {Battye}, R.~A., {et~al.} 2020, \pasa, 37, e007, \dodoi{10.1017/pasa.2019.51}

\bibitem[{{Staveley-Smith} \& {Oosterloo}(2015)}]{2015aska.confE.167S}
{Staveley-Smith}, L., \& {Oosterloo}, T. 2015, in Advancing Astrophysics with
  the Square Kilometre Array (AASKA14), 167, \dodoi{10.22323/1.215.0167}

\bibitem[{{Wolfe}(1988)}]{1988qsal.proc..297W}
{Wolfe}, A.~M. 1988, in Proceedings of the QSO Absorption Line Meeting, ed.
  J.~C. {Blades}, D.~A. {Turnshek}, \& C.~A. {Norman}, 297--306

\bibitem[{{Yahya} {et~al.}(2015){Yahya}, {Bull}, {Santos}, {Silva}, {Maartens},
  {Okouma}, \& {Bassett}}]{2015MNRAS.450.2251Y}
{Yahya}, S., {Bull}, P., {Santos}, M.~G., {et~al.} 2015, \mnras, 450, 2251,
  \dodoi{10.1093/mnras/stv695}

\bibitem[{{Yoon} {et~al.}(2024){Yoon}, {Sadler}, {Mahony}, {Aditya}, {Allison},
  {Glowacki}, {Kerrison}, {Moss}, {Su}, {Weng}, {Whiting}, {Wong},
  {Callingham}, {Curran}, {Darling}, {Edge}, {Ellison}, {Emig},
  {Garratt-Smithson}, {German}, {Grasha}, {Koribalski}, {Morganti},
  {Oosterloo}, {P{\'e}roux}, {Pettini}, {Pimbblet}, {Zheng}, {Zwaan}, {Ball},
  {Bock}, {Brodrick}, {Bunton}, {Cooray}, {Edwards}, {Hayman}, {Hotan},
  {Lee-Waddell}, {McClure-Griffiths}, {Ng}, {Phillips}, {Raja}, {Voronkov}, \&
  {Westmeier}}]{2024arXiv240806626Y}
{Yoon}, H., {Sadler}, E.~M., {Mahony}, E.~K., {et~al.} 2024, arXiv e-prints,
  arXiv:2408.06626, \dodoi{10.48550/arXiv.2408.06626}

\bibitem[{Yu {et~al.}(2014)Yu, Zhang, \& Pen}]{Yu_2014}
Yu, H.-R., Zhang, T.-J., \& Pen, U.-L. 2014, Physical Review Letters, 113,
  \dodoi{10.1103/physrevlett.113.041303}

\end{thebibliography}
\bibliographystyle{aasjournal}
\end{document}